\documentstyle[12pt,aaspp4,epsf]{article}
\slugcomment{To Appear the in {\it The Astrophysical Journal}}

\lefthead{Cavallo, Sweigart, \& Bell}
\righthead{Nucleosynthesis in Globular Cluster Red Giant Stars}
\begin{document}

\title{Proton-Capture Nucleosynthesis in Globular Cluster Red Giant Stars}
\author{Robert M. Cavallo}
\affil{Department of Astronomy, University of Maryland, College Park, MD
20742; rob@astro.umd.edu}
\authoraddr{Department of Astronomy, University of Maryland, College Park, MD
20742; rob@astro.umd.edu}
\author{Allen V. Sweigart}
\affil{Laboratory for Astronomy and Solar Physics, Code 681, NASA/Goddard
Space Flight Center, Greenbelt, MD 20771; sweigart@bach.gsfc.nasa.gov}
\authoraddr{Laboratory for Astronomy and Solar Physics, Code 681, NASA/Goddard
Space Flight Center, Greenbelt, MD 20771; sweigart@bach.gsfc.nasa.gov}
\and
\author{Roger A. Bell}
\affil{Department of Astronomy, University of Maryland, College Park, MD
20742; roger@astro.umd.edu}
\authoraddr{Department of Astronomy, University of Maryland, College Park, MD
20742; roger@astro.umd.edu}

\begin{abstract}

Observational evidence suggests that many of the variations of the surface
 abundances of light to intermediate mass elements (A $<$ 28) in globular
 cluster red-giant-branch (RGB) stars can be attributed to non-canonical mixing
 between the surface and the deep stellar interior during the RGB phase.
As a first step to studying this mixing in more detail, we have combined
 a large nuclear reaction network with four detailed stellar evolutionary 
 sequences of different metallicities in order to follow the production and
 destruction of the C, N, O, Ne, Na, Mg, and Al isotopes around the
 hydrogen-burning shell (H shell) of globular cluster RGB stars.
The abundance distributions determined by this method allow for 
 the variation in the temperature and density around the H shell as well as
 for the dependence on both the stellar luminosity and cluster metallicity.
Because our nuclear reaction network operates separately from the stellar
 evolution code, we are able to more readily to explore the effects of the
 uncertainties in the reaction rates on the calculated abundances.

We discuss implications of our results for mixing in the context of
 the observational data.
Our results are qualitatively consistent with the observed C vs. N, O vs. N,
 Na vs. O, and Al vs. O  anticorrelations and their variations with both
 luminosity and metallicity.
We see evidence for variations in Na without requiring changes in O, independent
 of metallicity, as observed by Norris \& Da Costa (\markcite{r46}1995a)
 and Briley et al.  (\markcite{r122}1997).
Also, we derive $^{12}$C/$^{13}$C ratios near the observed equilibrium value
 of 4 for all sequences, and predict the temperature-dependent
 $^{16}$O/$^{17}$O equilibrium ratio based on new data for the
 $^{17}$O$({\it p,{\alpha}})^{14}$N reaction rate.
Additionally, we discuss the Mg isotopic abundances in light of the recent
 observations of M13 (Shetrone \markcite{r116}1996b) and NGC 6752 (Shetrone
 \markcite{r124}1997).
\end{abstract}


\section{Introduction}

Stars which lie near each other on the color-magnitude diagrams of
 globular clusters were initially thought to have similar chemical compositions.
However, observations over the past two decades have shown a
 large scatter of the abundances of C, N, O, Na, Mg, and Al among cluster
 red-giant stars [for reviews, see Kraft \markcite{r1}(1994) and Briley et al.
 \markcite{r2}(1994)].
Explanations for these abundance variations have generally focused on two
 possibilities: 1) the mixing of nucleosynthesized material from the deep 
 stellar interior during the red-giant-branch (RGB) evolution, and 2)
 primordial abundance variations within the globular cluster material.
In this work we model those abundance variations which can be attributed to
 mixing during the red-giant-branch (RGB) evolution by using a detailed 
 nuclear reaction network together with four RGB stellar evolutionary
 sequences of different metallicities.
We make no attempt to model those variations which are likely the
 result of primordial contamination from prior stellar evolution, such as
 the observed Fe-peak metallicity spread in ${\omega}$ Cen (Dickens \& Bell
 \markcite{r11}1976), the Ca abundance variations in M22 (Anthony-Twarog,
 Twarog, \& Craig \markcite{r13}1995), and the Na variations on the upper
 main sequence of 47 Tuc (Briley et al. \markcite{r89}1996).

The decrease in C with increasing luminosity seen in the very
 metal-poor ([Fe/H]
 \footnote{This follows the notation: [X] = log(X)${_\star} -$
 log(X)${_\odot}$} ${\lesssim} -2.0$)
 and intermediate metallicity ($-2.0 {\lesssim}$ [Fe/H] ${\lesssim} -1.1$)
 clusters supports the mixing hypothesis.
In the intermediate-metallicity and metal-rich ([Fe/H] $\gtrsim -$1.1)
 clusters where $^{13}$C is also observable, the $^{12}$C/$^{13}$C ratio is 
  seen to decrease to near its equilibrium value as the luminosity increases,
  providing further evidence of mixing on the RGB.
 (Bell, Dickens, \& Gustafsson \markcite{r3}1979; Bell \& Dickens
 \markcite{r8}1980; Da Costa
 \& Cottrell \markcite{r57}1980; Carbon et al. \markcite{r4}1982; Trefzger et 
 al. \markcite{r7}1983; Langer et al. \markcite{r5}1986; Smith \& Suntzeff
 \markcite{r52}1989; Briley et al. \markcite{r6}1990; Suntzeff \& Smith
 \markcite{r14}1991; Brown, Wallerstein, \& Oke \markcite{r55}1991).
Along with these changes, observations of the CN and NH bands have shown that C
 is anticorrelated with N at all metallicities (Hesser, Hartwick, \& McClure
 \markcite{r48}1977; Norris \& Freeman \markcite{r31}1979; Norris et al.
 \markcite{r20}1981; Suntzeff \markcite{r9}1981; Langer, Kraft, \& Friel
 \markcite{r25}1985; Smith \& Bell \markcite{r26}1986).
Furthermore, N and Na are both anticorrelated with O and their abundances
 are seen to increase 
 with increasing luminosity (Cottrell \& Da Costa \markcite{r76}1981; 
 Pilachowski \markcite{r64}1988; Paltoglou \& Norris \markcite{r43}1989;
 Dickens et al. \markcite{r78}1991; Sneden et al. \markcite{r65}1991; 
 Drake, Smith, \& Suntzeff \markcite{r84}1992; Sneden et al.
 \markcite{r85}1992; Norris \& Da Costa \markcite{r46}1995a; Smith et al.
 \markcite{r69}1996; Pilachowski et al. \markcite{r90}1996; hereafter, PSKL96).
The variations in Na, first found by Cohen \markcite{r74}(1978) and
 Peterson \markcite{r75}(1980), were not initially thought to be the result of 
 internal stellar evolution; however, Denisenkov \& Denisenkova 
 \markcite{r93}(1990) and Langer, Hoffman, \& Sneden \markcite{r94}(1993) 
 showed that Na could be produced from proton captures on Ne at the same
 temperatures that O is destroyed.

The Al abundance also shows an anticorrelation with O, but with a stronger
 dependence on metallicity than Na (Norris \& Da Costa \markcite{r46}1995a;
 \markcite{r86}1995b); some metal-poor clusters exhibit large Al
 enhancements by as much as 1.2 dex, while
 the more metal-rich clusters have no Al variations (Cottrell \& Da Costa
 \markcite{r76} 1981).
Langer, Hoffman, and Sneden (\markcite{r94}1993) showed that Al can be
 produced from proton captures on $^{25,26}$Mg, and subsequently,
Langer \& Hoffman (\markcite{r95}1995) attempted to explain the large Al 
 overabundances in M13 (Kraft et al. \markcite{r113}1997) by suggesting that
 the $^{25,26}$Mg isotopes were initially overabundant by a factor of 2.
However, the observations of Mg in M13 by PSKL96 and Shetrone
 (\markcite{r91}1995) do not support this conclusion.

The observations of the Mg abundances show the most erratic trends.
PSKL96 showed that fainter red giants with small ($<$ 0.10 dex) Na 
 enhancements had large Mg abundances.
However, while roughly half of the giants with large Na enhancements had
 low Mg, the rest were widely distributed in Mg abundance, by $\sim$ 0.6 dex.
For the high luminosity giants, most were Na-enriched while, on the other
 hand, more than half showed no evidence of Mg depletion.
The difficulty in describing the Mg abundance observations stems from the
 fact the there are three stable Mg isotopes which are relatively plentiful;
 e.g., log(N(Mg)/N(H))$_\odot$ = $-$4.42 with the
 $^{24}$Mg:$^{25}$Mg:$^{26}$Mg ratio
 equal to 79:10:11 on earth and presumably in the solar photosphere
 (Anders \& Grevesse \markcite{r104}1989).
Shetrone (\markcite{r114}1996a,\markcite{r116}1996b) argued that the change
 in the total Mg abundance for M13 giants near the RGB tip is controlled by
 $^{24}$Mg by demonstrating that $^{24}$Mg is anticorrelated with Al.
Other studies have shown total Mg vs. CN anticorrelations in ${\omega}$ Cen
 (Norris \& Da Costa \markcite{r46}1995a) and NGC 6752 
 (Shetrone \markcite{r124}1997), while still other studies found
 no star-to-star variations in Mg in NGC 6752, 47 Tuc, or M22
 (Hesser \markcite{r60}1978; Norris et al.\markcite{r20} 1981; Smith \& Wirth
 \markcite{r83}1991; Briley, Hesser, \& Bell \markcite{r63}1991).
In this work, we look at how the changes in these elemental abundances
 could be brought about during a star's ascent up the RGB.

There have been some theoretical attempts to explain these abundance variations.
For example, Sweigart \& Mengel (\markcite{r92}1979; hereafter, SM79) 
 showed that there were regions above the hydrogen-burning shell (H shell)
 where C and O were depleted.
They suggested that rotationally driven meridional circulation on the RGB could
 transport material from these depleted regions to the outer convective 
 envelope.
Their work qualitatively accounted for the C and O vs. N anticorrelations 
 as well as the decline in C and the $^{12}$C/$^{13}$C ratio with 
 increasing luminosity.
Although they used detailed stellar models of different metallicities,
 they lacked a full reaction network to study the variations in the elements
 beyond O and did not examine the uncertainties in the nuclear reaction 
 rates.

Denissenkov \& Weiss (\markcite{r98}1996) recently examined deep diffusive
 mixing in globular cluster and halo giants by constructing a very metal-poor
 (Z = 0.0004) stellar sequence containing four models and combining it with the
 widely used rates of Caughlin \& Fowler (\markcite{r97}1988; hereafter, CF88).
They used a two-parameter mixing scenario; one which controlled the
 mixing depth and the other which controlled the mixing rate.
They were able to model the observed changes in the $^{12}$C/$^{13}$C ratio
 and [C/Fe] with M$_{\rm V}$ in M92 as well as the global O vs. Na 
 anticorrelation observed in other clusters such as M3 and M13 (Kraft et al.
 \markcite{r68}1992, \markcite{r77}1993; hereafter, KSLP92 and KSLS93,
 respectively).
However, each correlation required a different set of parameters,
 and they were still unable to reproduce the large enhancement of $^{27}$Al
 seen in some clusters without increasing the initial $^{25,26}$Mg abundances 
 beyond the scaled solar value.
These difficulties may be due to the fact that they only considered
 a single sequence and were therefore unable to discuss metallicity-dependent 
 variations.
Furthermore, their use of the CF88 reaction rates did not allow them to
 take advantage of the updated data for the NeNa and MgAl cycles, which have 
 changed substantially since the work of CF88.

In our previous study (Cavallo, Sweigart, \& Bell \markcite{r96}1996;
 hereafter, Paper I), we used an extensive nuclear reaction network to
 compute the isotopic abundances around the H shell during the RGB evolution
 of two stellar model sequences of different metallicities.
The conclusions from Paper I were that $^{23}$Na is produced first by proton
 captures on $^{22}$Ne in the region where O begins to deplete, then from 
 $^{20}$Ne via the NeNa cycle deeper in the region of O depletion.
Further, $^{27}$Al is produced in the O-depleted region by proton captures,
 first on $^{25,26}$Mg, then on $^{24}$Mg in the MgAl cycle (see Fig. 1 of
 Paper I for a graphical description of the NeNa and MgAl cycles).
However, the replenishment of $^{24}$Mg by proton captures on $^{23}$Na
 occurred at a faster rate than its depletion to form $^{27}$Al.
Thus, while the results were in qualitative agreement with the observations 
 for Na and Al, they could not predict the observed Mg variations.

In this paper, we extend the work of Paper I by examining more fully the 
 effects of the nuclear reaction rates and stellar structure on the
 abundance profiles of C, N, O, Na, Mg, and Al around the H shell
 along four RGB stellar evolutionary sequences.
Our approach will include the effects of the variation in the temperature
 and density around the H shell on the isotopic abundance yields.
In addition, we will follow the changes in the abundances as the H shell
 evolves.
Finally, by using a detailed nuclear reaction network that is separate from
 the stellar evolutionary code, we can explore the uncertainties in the 
 reaction rates easily and efficiently.
We make no attempt to incorporate mixing at this stage and concentrate, instead,
 on the development of the abundance profiles in the sequences.

In section 2 we describe the computational procedure used both here
 and in Paper I.
We discuss the role of temperature and density in controlling the shape and
 lifetime of the abundance profiles around the H shell in section 3.
Our results for the abundance profiles are presented in section 4, while in
 section 5 we examine these results in the context of mixing.
In section 6 we explore the impact of the uncertainties in the nuclear reaction
 rates on our results, and in section 7 we summarize the observational
 implications of our results.

\section{Computational Procedure}
\subsection{Stellar Evolutionary Sequences}

We have computed four stellar evolutionary sequences, three
 which cover the range of metallicities seen in the globular clusters and 
 one which corresponds to an old disk star with solar metallicity.
The masses of the globular cluster sequences are chosen to give 
 an age of $\sim$ 15 Gyr at the tip of the RGB.
Since the evolution up the RGB does not depend strongly on the mass of the 
 star, the choice of age does not significantly affect our results.
The sequences are begun at the zero-age main sequence and evolved up the RGB
 to the He flash at the tip of the RGB.
The input physics for these model computations is outlined in Sweigart
 (\markcite{r115}1997a).
The properties of the sequences are given in Table 1.
They contain from 1600 to 4000 models between the onset of mixing, as discussed
 below, and the He flash.
From these we have selected models at increments of 0.005 M$_\odot$ in 
 M$_{\rm sh}$, where M$_{\rm sh}$ is the mass interior to the H 
 shell, or, equivalently, the mass of the helium core.
Each model provides us with the run of temperature and density around
 the H shell as well as the hydrogen-shell luminosity, effective temperature, 
 and age.

\begin{deluxetable}{lllccccccrc}
\scriptsize
\tablewidth{0pt}
\tablecaption{Properties of the RGB Sequences}
%\tablecaption{Properties of the RGB Sequences \label{t1}}
\tablenum{1}
\tablecolumns{11}
\tablehead{
 \colhead{Z}                 &
 \colhead{[Fe/H]}            &
 \colhead{Y}                 &
 \colhead{M}                 &
 \colhead{Log $L_{\rm mix}\tablenotemark{a}$} &
 \colhead{M$_{\rm sh,mix}\tablenotemark{b}$}  &
 \colhead{Log $L_{\rm tip}\tablenotemark{c}$} &
 \colhead{M$_{\rm sh,tip}\tablenotemark{d}$}  &
 \colhead{$\tau_{\rm RGB}\tablenotemark{e}$}  &
 \colhead{$t_{\rm tip}\tablenotemark{f}$}     &
 \colhead{Models}            \\
 \multicolumn{3}{c}{}        &
 \colhead{(M$_\odot$)}       &
 \colhead{($L_\odot$)}        &
 \colhead{(M$_\odot$)}       &
 \colhead{($L_\odot$)}        &
 \colhead{(M$_\odot$)}       &
 \colhead{(Myr)}             &
 \colhead{(Gyr)}             &
 \colhead{Selected\tablenotemark{g}}          
}
\startdata
%   Z      [Fe/H]      Y       M   Lmix   Mshmix Lflash MshHe TauRGB AgeTip #models
0.0001  & $-$2.27  & 0.230  & 0.795 & 2.28 & 0.356 & 3.22 & 0.501 & 20.8 & 15.0  & 31 \nl
0.0004  & $-$1.67  & 0.230  & 0.800  & 2.11 & 0.325 & 3.26 & 0.492 & 29.1 & 15.1 & 36 \nl
0.004   & $-$0.67  & 0.230  & 0.875 & 1.72 & 0.273 & 3.33 & 0.483 & 58.7 & 14.9  & 44 \nl
0.01865 & {\phs}0.00  & 0.2825 & 1.500 & 1.83 & 0.272 & 3.34 & 0.464 & 40.5 & 2.7  & 40 \nl

\tablenotetext{a}{The luminosity at the start of mixing.}
\tablenotetext{b}{The mass coordinate M$_r$ at the center of the H shell at
		  the start of mixing.}
\tablenotetext{c}{The luminosity at the tip of the RGB.}
\tablenotetext{d}{The mass coordinate M$_r$ at the center of the H shell at
		  the tip of the RGB.}
\tablenotetext{e}{The lifetime on the RGB between the start of mixing and the
		  He flash.}
\tablenotetext{f}{The sequence age at the tip of the RGB.}
\tablenotetext{g}{The number of models from each sequence in which the
		  abundance profiles were calculated.}

\enddata
\end{deluxetable}

A fundamental question is to decide where along the RGB to begin the mixing
 process.
As discussed by SM79, the hydrogen discontinuity left behind by the deep
 penetration of the convective envelope during the first dredge-up probably
 prevents the circulation currents from accessing the regions of C-N and O-N
 conversion.
However, once the H shell has burned past this discontinuity, the currents
 are no longer inhibited by a mean molecular weight gradient, and hence, the
 region above the H shell can then be mixed.
This point along the RGB corresponds to the well-known bump in the
 RGB luminosity function (Fusi Pecci et al. \markcite{r117}1990).
This hypothesis appears to be supported by the work of Richard et al.
 (\markcite{r129}1996) who showed that there is a critical ${\mu}$-gradient
 in the sun which is sufficient to stop rotationally induced circulation
 of matter.
The results of Bell, Dickens, \& Gustafsson (\markcite{r3}1979) add 
 observational evidence to this assumption.
Gilroy \& Brown (\markcite{r17}1991) have demonstrated that in the old open
 cluster M67 ([Fe/H] $\sim$ 0) the $^{12}$C/$^{13}$C ratio does not begin
 to decline past its canonical value until the predicted start of mixing.
Likewise, Suntzeff (\markcite{r9}1981) notes that the abundances in M3 giants
 appear to follow the SM79 presumption, but those in M13 don't.
Furthermore, Charbonnel (\markcite{r130}1994, \markcite{r131}1995) has
 successfully modeled the observed variation of $^{12}$C/$^{13}$C, $^{7}$Li,
 and $^{12}$C/$^{14}$N in both globular and open clusters under this assumption.

There are, however, observations which show that abundance anomalies persist at
 luminosities below the point at where the H shell burns through the hydrogen
 discontinuity.
In addition to the previously noted Na variations seen in 47 Tuc, some
 observers have seen star-to-star differences in the CN content of lower
 RGB and sub-giant-branch (SGB) stars in the clusters M3, M4, M5, M92,
 NGC 6752, and 47 Tuc (Bell, Hesser, \& Cannon \markcite{r62}1983; Norris 
 \& Smith \markcite{r24}1984; Langer et al. \markcite{r5}1986; Smith, Bell, \&
 Hesser \markcite{r33}1989; Suntzeff \& Smith \markcite{r14}1991; 
 Briley et al. \markcite{r28}1992).
In these cases, the CH band strength appears to be weakly anticorrelated with
 the CN band strength.
However, at these lower luminosities, the G band of CH is difficult to
 measure with the necessary accuracy to ascertain the CH abundance
 (Bell, Hesser, \& Cannon \markcite{r34}1984).
As pointed out by VandenBerg \& Smith (\markcite{r135}1988), the best hope
 of diagnosing the inherent nature of the CN anomalies (primordial or mixing)
 is to measure the $^{12}$C/$^{13}$C ratio, which drops sharply at the onset
 of mixing.
At present, there is a paucity of data concerning this ratio at such low
 luminosities.

In this work we will assume that mixing begins at the point along the RGB
 where the H shell burns through the hydrogen discontinuity and save
 an investigation into mixing on the lower RGB and SGB for another project.

\subsection{Nuclear Reaction Network}
Our nuclear reaction network was kindly supplied by Dr. David Arnett
 of the University of Arizona.
The original version relied mostly on the rates of CF88;
 we have since modified the code to reflect more modern results for the rates 
 in the CNO, NeNa, and MgAl cycles at the temperatures around the H
 shell of our models.
We discuss the reaction rates in detail and examine the effects of their
 uncertainties on our results in section 6.

This network code requires three input ingredients: the run of temperature and
 density around the H shell, the initial envelope composition, and the 
 timescale for the nuclear reactions.
The last ingredient is obtained by the stationary shell approximation which is
 discussed in the following subsection.
As indicated above, we obtain the run of temperature and density from the
 RGB models.

We supply an initial composition for all isotopes from F through Ca according
 to their scaled solar abundance (Anders \& Grevesse \markcite{r104}1989).
The envelope values of the C, N, and O isotopes are given by the models
 and include the effects of the first dredge-up at the base of the RGB.
The network follows the reactions of 113 nuclei from $^{1}$H to $^{49}$Ca
 and produces the abundance profiles in the region of the H shell for
 each model.

\subsection{Stationary Shell Approximation}

There are two ways to compute the distribution of the elements around the
 H shell.
The most direct method is to incorporate the nuclear reaction network into
 the stellar evolution code.
This becomes very expensive computationally, especially when trying to
 to explore the uncertainties in the various reaction rates.
A simpler method is to derive the models first and then to compute the 
 composition distribution afterwards via the stationary shell
 approximation (SSA), which we outline here.
In addition to being more computationally efficient, the SSA also allows us to 
 vary the reaction rates, and (later) the mixing algorithm parameters,
 without having to recompute each evolutionary sequence.

In order to integrate the network code, one must know the timescale  of
 the nuclear burning.
This burning time can be straightforwardly determined through the use of the
 SSA, as we now demonstrate.
Figure 1 shows the H-shell profile at two different times, t$_1$ and t$_2$.
\begin{figure}
\plotone{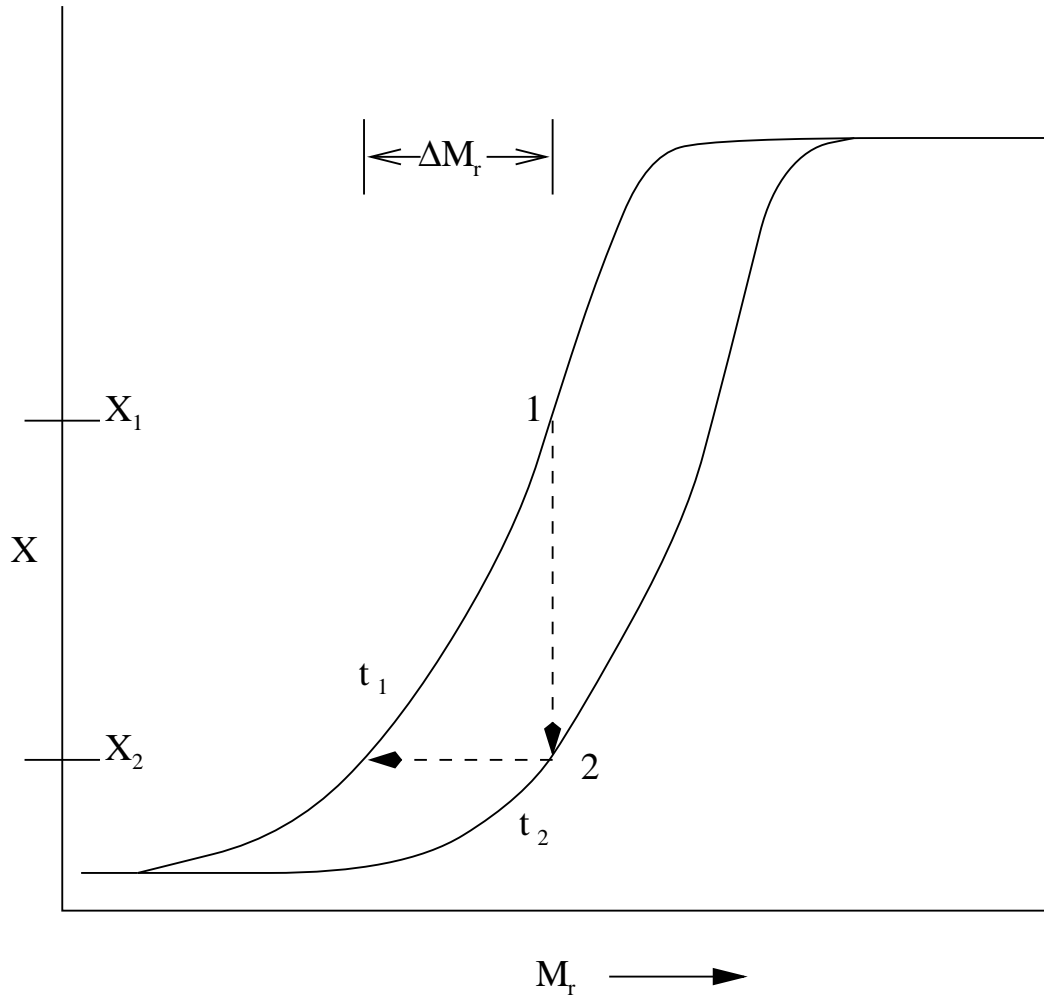}
\caption{Schematic diagram of the H shell at two different
 times t$_1$ and t$_2$.}
\end{figure}
At an arbitrary point in the shell the hydrogen abundance decreases from 
 X$_1$ at time t$_1$, to X$_2$ at time t$_2$ according to the relation:
\begin{equation}
  \eqnum{1}
  \Delta {\rm X = X_1 - X_2 = \frac{dX}{dt}\Delta t,}
\end{equation}
 where $\Delta {\rm t}$ = t$_2$ - t$_1$ is the burning time.
{\em Provided that the shell profile does not change significantly over
 the time interval} $\Delta {\rm t}$, we can also compute the change in X from
  the equation:
\begin{equation}
  \eqnum{2}
  \Delta {\rm X = \frac{dX}{dM_r}\Delta M_r,}
\end{equation}
 where ${\Delta}$M$_{\rm r}$ is the amount of mass through which the H shell
 advances over the time interval ${\Delta}$t.
Thus, changing from X$_1$ to X$_2$ can be accomplished either by remaining 
 at a fixed value of the mass coordinate M$_{\rm r}$ and allowing the shell
 to advance outward over a time $\Delta$t,
 or by integrating along the profile over the mass interval $\Delta {\rm M_r}$.

The next step is to relate ${\Delta}$t to $\Delta$M$_{\rm r}$.
Letting X$_{\rm e}$ denote the envelope hydrogen abundance, it follows that the
 mass of hydrogen, $\Delta$M$_{\rm X}$, burned over the time interval $\Delta$t
 can be written as
\begin{displaymath}
 \Delta {\rm M_X = (\Delta M_r)X_e.}
\end{displaymath}
Since the energy liberated by burning this mass of hydrogen must be equal to 
 the total energy output of the shell over the time interval $\Delta$t, we can
 write 
\begin{equation}
 \eqnum{3}
 {\rm (\Delta M_r)X_eE = L_{H}{\Delta}t},
\end{equation}
 where E is the energy released in burning one gram of hydrogen and L$_{\rm H}$
 is the hydrogen shell luminosity.
Combining equations (1) - (3) yields the desired relationship 
\begin{equation}
 \eqnum{4}
 {\rm \frac{dX}{dt} = \frac{L_H}{X_eE}\frac{dX}{dM_{r}}}.
\end{equation}

Using equation (4), we can convert the time derivative of the hydrogen 
 abundance, or any other isotopic abundance, as given by the nuclear reaction
 rates, to the corresponding derivative with respect to M$_{r}$.
From this we can then determine the abundance distribution of the all of the 
 isotopes around the H shell.

How well does the SSA work?
In Figure 2 we compare the H and CNO abundances found with the SSA for
 a representative RGB model with the abundances computed with the stellar
 evolutionary code (see, e.g., Clayton \markcite{r103}1983).
The agreement is excellent at all points and justifies the use of the SSA 
 to set the nuclear burning time.

\begin{figure}
\plotone{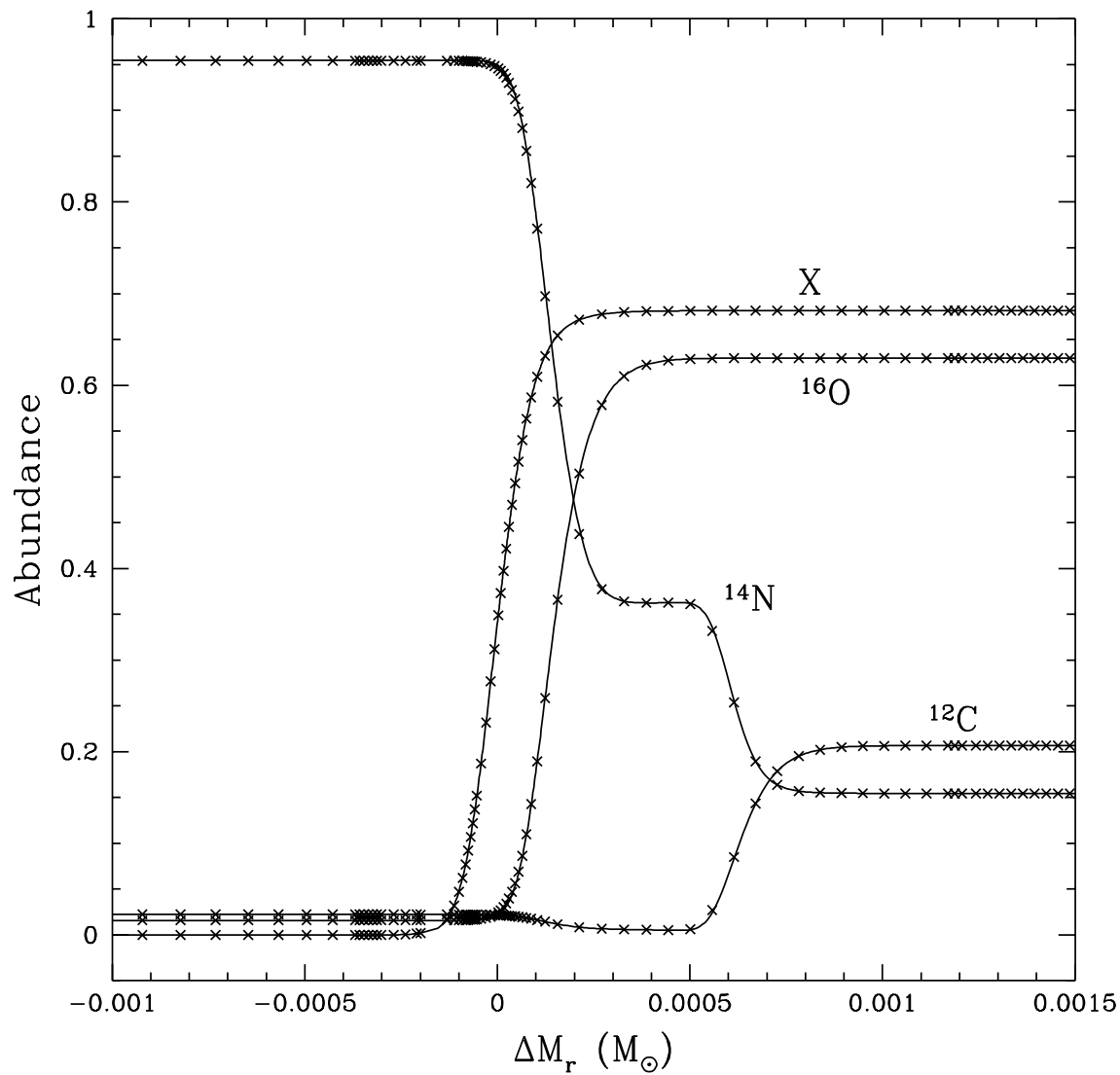}
\caption{Comparison between the H and CNO abundances
 computed with the stationary shell approximation (x's) and the stellar code
 (solid line) for a representative RGB model. The ordinate
 is the number abundance relative to all metals.  The curve labeled by
 X is the hydrogen-mass fraction on the same scale. The abscissa is the
 mass difference between any point and the center of the H shell.}
\end{figure}

\section{Properties of the H shell}

The isotopic abundance profiles around the H shell depend on the run of
 temperature and density and on the timescale, ${\tau}_{\rm shell}$, with which
 the shell advances outward in mass.
In this section we discuss each of these in turn.

\subsection{Run of T and ${\rho}$ around the H shell}

In Figure 3a we plot the temperature profiles around the H shell for
 the Z = 0.0004 sequence at four values of M$_{\rm sh}$.
\begin{figure}
\plotone{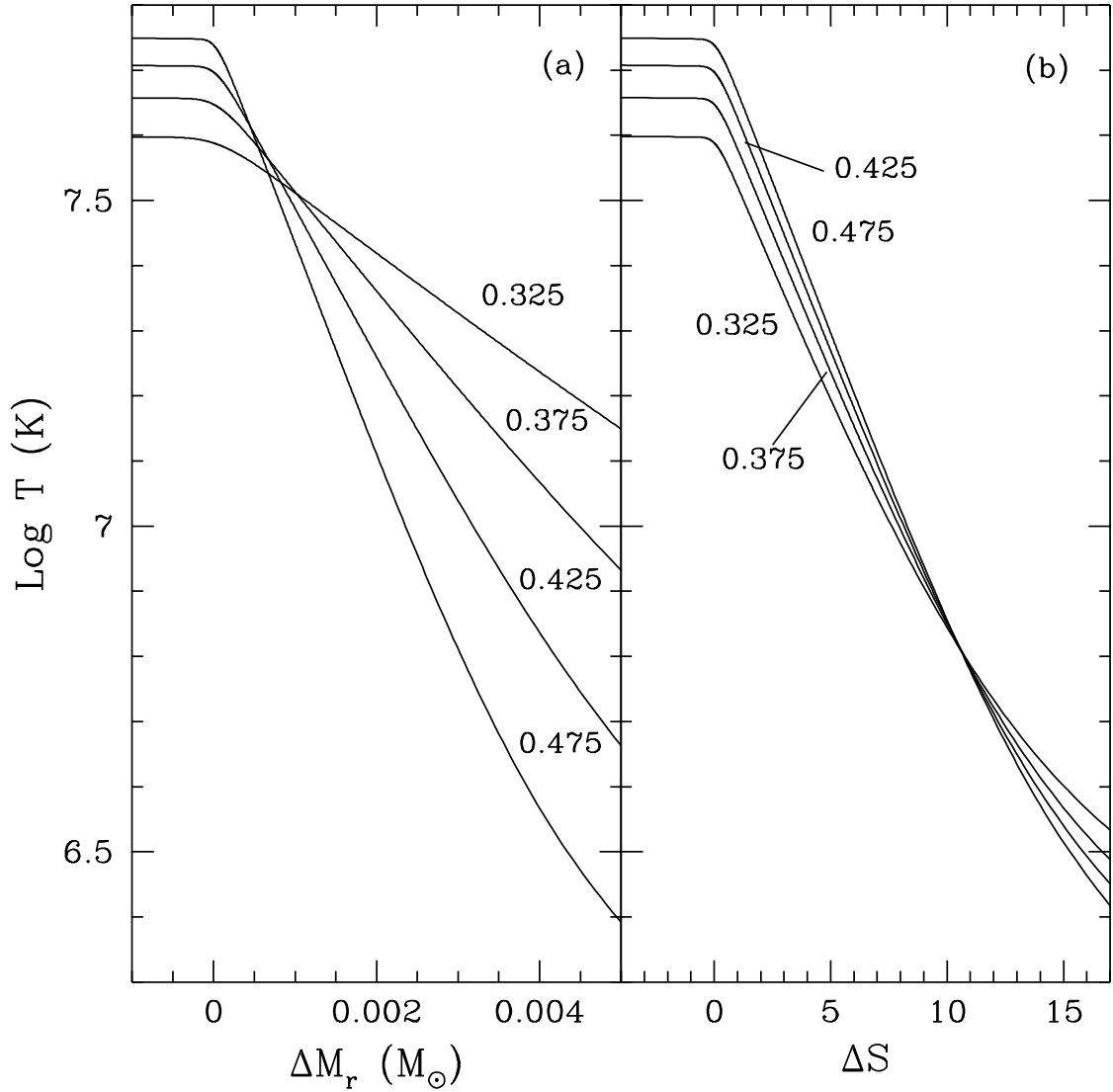}
\caption{(a) Temperature profiles around the H shell as
 a function of the mass difference ${\Delta}$M$_{r}$ between any point and the
 center of the H shell for four models from the Z = 0.0004 sequence.
Each curve is labeled by the mass, M$_{\rm sh}$, interior to the H shell.
(b) The same profiles, except as a function of ${\Delta}$S.}
\end{figure}
In order to explore various mixing algorithms, one would like to be able to
 interpolate accurately between these temperature profiles for an arbitrary
 point along the RGB.
However, interpolating between these profiles is difficult due to their 
 overlapping just above the H shell.
In order to remedy this problem, we introduce the variable ${\Delta}$S,
 defined by
\begin{displaymath}
  \eqnum{5}
  {\Delta} {\rm S = {\Delta}M_{\rm r}/{\Delta}{\rm M_{H}},}
\end{displaymath}
 where ${\Delta}$M$_{\rm r}$ = M$_{\rm r}- $M$_{\rm sh}$ is the mass difference
 between any point and the center of the H shell, and ${\Delta}$M$_{\rm H}$ is
 the thickness of the H shell, defined as the mass between the points
 where the H-mass fraction, X, is 10\% (${\Delta}$S = $-$0.5) and 90\% 
 (${\Delta}$S = $+$0.5) of its envelope value.
Panel (b) of Figure 3 demonstrates that using ${\Delta}$S keeps the temperature
 curves separated near the H shell.

In Figure 4 we demonstrate the full range of temperatures around the H shell
 for all four sequences by plotting $T_9$ (= $T/10^9$K) as a function of 
 ${\Delta}$S at the start of mixing (top panel) and at the onset of the 
 helium flash (bottom panel).
\begin{figure}
\plotone{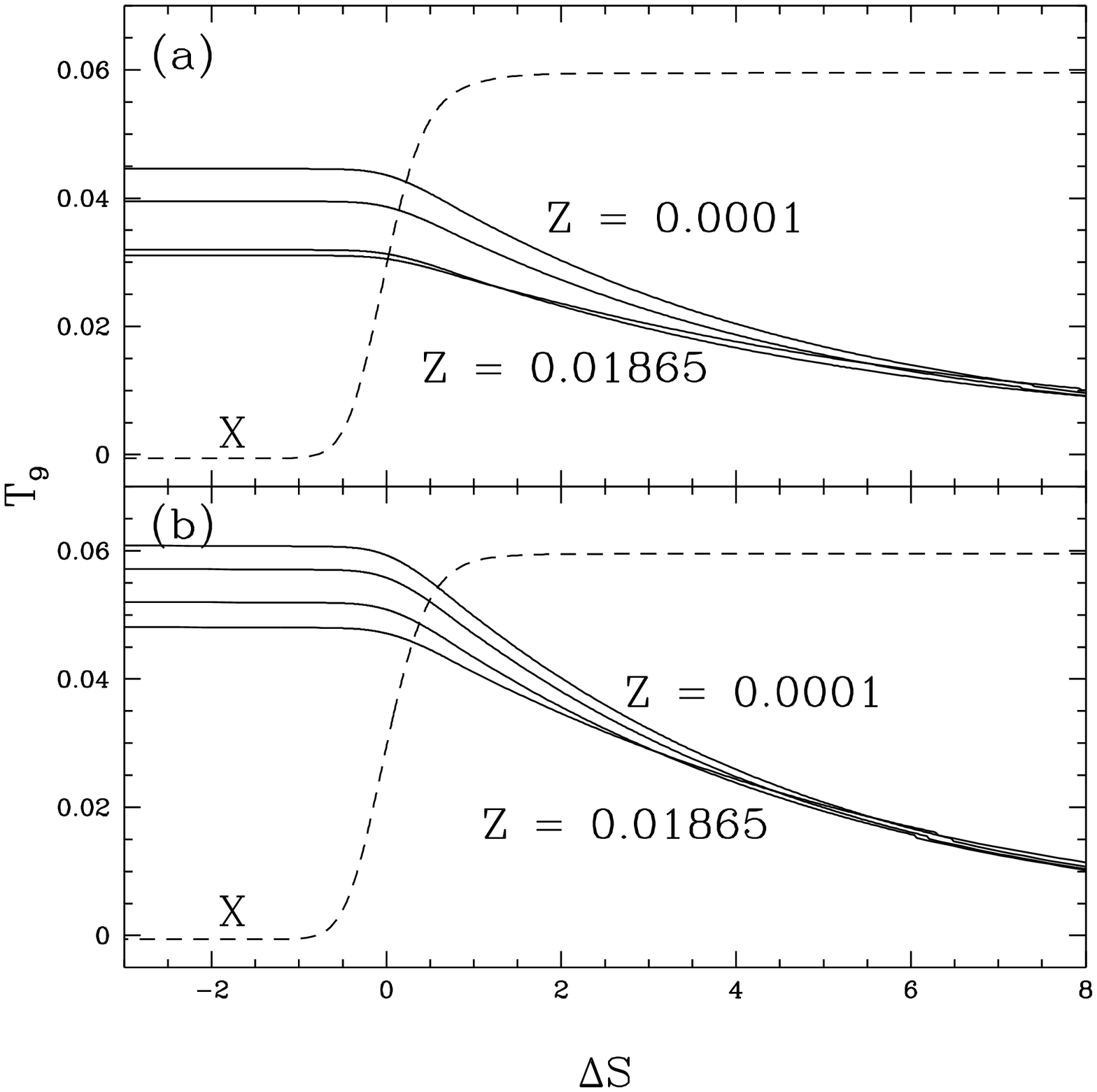}
\caption{The range in temperature, $T_9$, around the H
 shell for all four sequences at (a) the assumed onset of mixing and (b) the
 tip of the RGB.  The dashed line is the hydrogen-mass fraction on an
 arbitrary scale.}
\end{figure}
The higher shell temperatures toward the tip of the RGB allow the occurrence
 of some nuclear reactions that were not possible lower down the RGB.
In addition, the metal-poor sequences reach higher temperatures in
 the shell and allow some elements to undergo nuclear processing 
 that would otherwise not be burned in the higher metallicity stars.

The density around the H shell for the same models plotted in Figure 3 is
 shown in Figures 5a and 5b.
Again, by using ${\Delta}$S we find that the density profiles near the H shell
 vary much more smoothly during the evolution.
\begin{figure}
\plotone{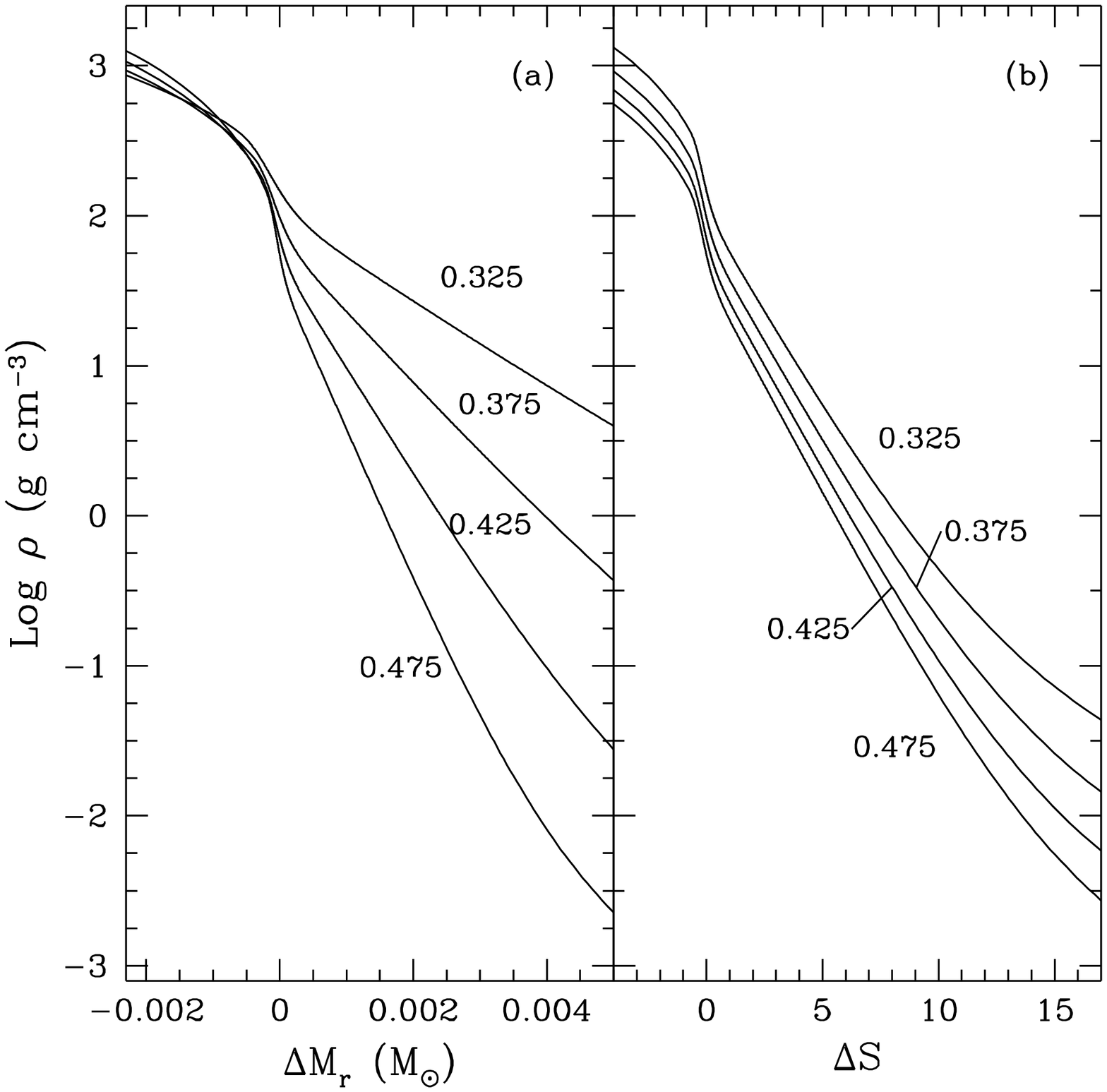}
\caption{As Figure 3, but for the density.}
\end{figure}
Although the density varies over many orders of magnitude near the H shell,
 its influence on the burning rates around the shell is only linear and,
 therefore, much less significant than the temperature dependence.
The density typically ranges from $\sim$ 0.01 to 1000 g cm$^{-3}$ for all
 four sequences.

\subsection{${\tau}_{\rm shell}$}

In addition to their dependence on temperature and density, the composition
 profiles also depend on the timescale with which the shell burns outward in
 mass.
We call this quantity ${\tau}_{\rm shell}$ and define it as the time required
 for the shell to move through one shell thickness, ${\Delta}{\rm M_H}$.
The dependence of this timescale on luminosity is given in
 Figure 6a for the portion of each sequence where mixing is presumed to occur.
\begin{figure}
\plotone{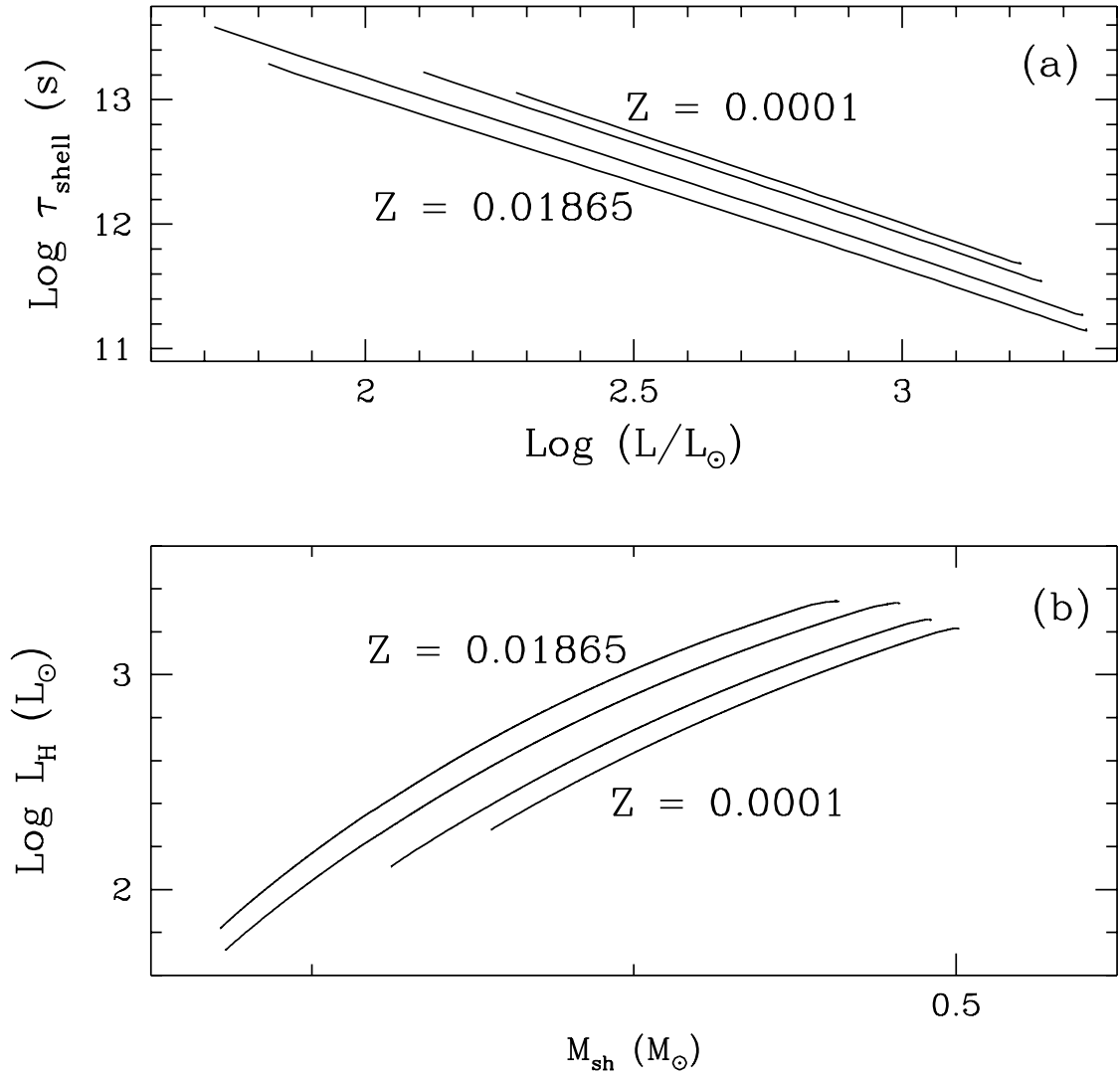}
\caption{(a) The shell burning timescale, ${\tau}_{\rm shell}$, as a function
 of luminosity for the four stellar evolutionary sequences.
 (b) The hydrogen-burning luminosity as a function of the mass, M$_{\rm sh}$,
 interior to the H shell for the same sequences.}
\end{figure}
The corresponding dependence of the hydrogen-burning luminosity, L$_{\rm H}$,
 on the core mass, M$_{\rm sh}$, is shown in Figure 6b.
We see from this panel that at a given core mass, the higher metallicity
 sequences have a greater hydrogen burning luminosity due to a larger number
 of CNO nuclei.
This increase in luminosity with metallicity, together with the increase in 
 the luminosity with evolution up the RGB, translates into a
 shorter ${\tau}_{\rm shell}$, as shown explicitly in panel (a).
Thus the higher metallicity and more evolved stars have less time to set up
 the composition profiles above the H shell.

The shell-burning timescale also depends on the shell thickness 
 ${\Delta}$M$_{\rm H}$, which is plotted in Figure 7 as a function of 
 M$_{\rm sh}$ for each metallicity.
As a sequence evolves, the increase in the luminosity causes the temperature 
 gradient across the H shell to steepen.
As a consequence the shell burning region narrows with evolution.
Moreover, the more metal-rich stars have steeper temperature gradients due
 to their larger opacities and greater flux, and thus, have narrower
 H shells at a given value of M$_{\rm sh}$.
Therefore, at brighter luminosities and higher metallicities,
 one expects the nuclear processing to occur in narrower burning regions,
 as implied by Figure 7.
\begin{figure}
\plotone{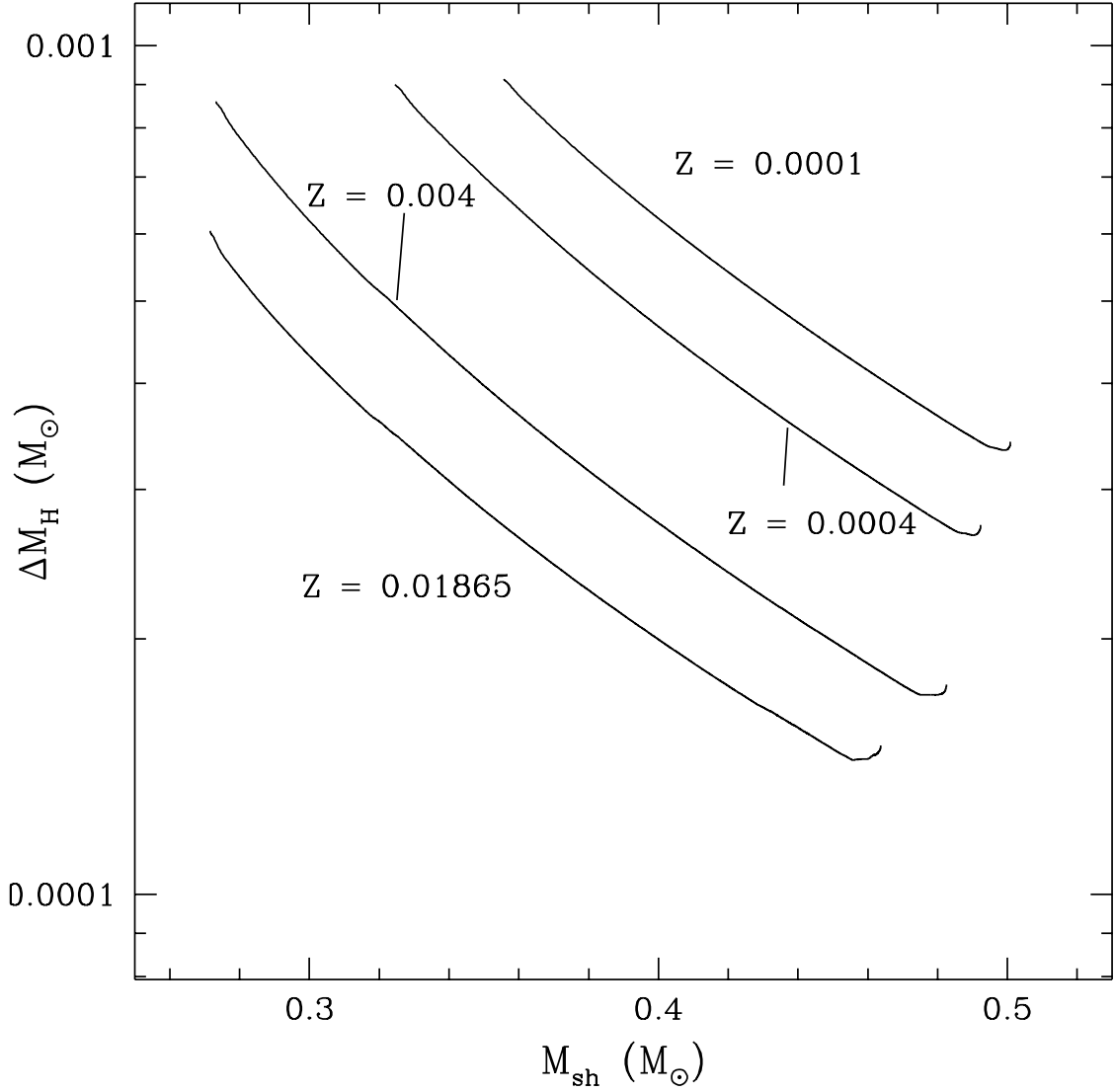}
\caption{The shell thickness, ${\Delta}{\rm M_H}$, as a
 function of the mass, M$_{\rm sh}$, interior to the H shell for the four
 evolutionary sequences.  The small upturn near the tip of the RGB is due to
 the onset of the helium flash.}
\end{figure}

\section{Isotopic Abundance Profiles}

The nuclear reaction network code produces the abundance profiles around the
 H shell for any isotope followed in the network.
Here we discuss the profiles of the CNO-cycle elements first, then the NeNa-
 and MgAl-cycle elements.

\subsection{C-N-O}

Figures 8 - 11 give the C, N, and O profiles around the H shell of our
 sequences, at the start of mixing (top panels) and at the tip of the
 RGB (bottom panels).
Above the H shell in each sequence is a region of C depletion in which 
 the temperature is hot enough only to convert $^{12}$C to $^{14}$N in the
 CN cycle. 
Closer in toward the H shell there exists an O-depleted region where 
 $^{16}$O is converted into $^{14}$N in the ON cycle.
The narrowing of the C- and O- depleted regions with increasing luminosity
 is seen between the top and bottom panels in each of these figures.
For example, at the onset of mixing in the Z = 0.0004 sequence (Figure 9a), 
 C depletion occurs at ${\Delta}$M$_{\rm r} = $ 0.0035 M$_{\odot}$.
As the sequence evolves to the tip of the RGB, $^{12}$C begins to be depleted
 much closer to the H shell, near ${\Delta}$M$_{\rm r} =$ 0.001 M$_{\odot}$. 
A similar narrowing of the C- and O-depleted regions occurs when the 
 metallicity increases.
For example, the Z = 0.0004 sequence at the start of mixing (Figure 9a) 
 experiences a depletion of $^{16}$O around ${\Delta}$M$_{\rm r} =$ 0.0015
 M$_{\odot}$ while the Z = Z$_\odot$ sequence (Figure 11a) depletes $^{16}$O
 much closer to the H shell at ${\Delta}$M$_{\rm r} =$ 0.0005 M$_{\odot}$.

\begin{figure}
\plotone{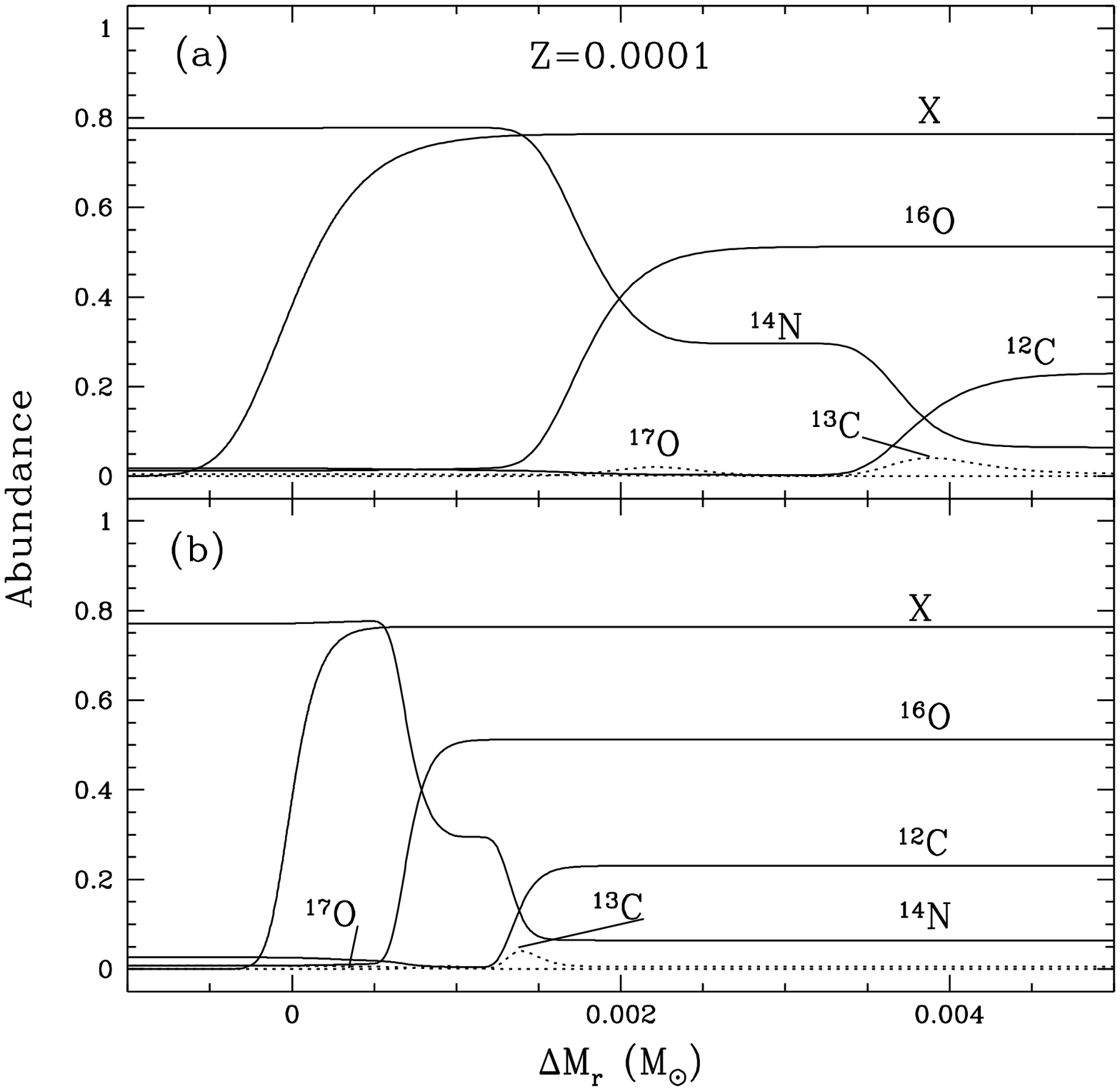}
\caption{The CNO isotopic abundance profiles around the H
 shell (labeled by the H-mass fraction X) for the Z = 0.0001 sequence at (a)
 the presumed onset of mixing, and (b) the tip of the RGB. The ordinate
 is the number abundance relative to all metals. The abscissa is the mass 
 difference, ${\Delta}$M$_{\rm r}$, between any point and the center of the
 H shell.}
\end{figure}
\begin{figure}
\plotone{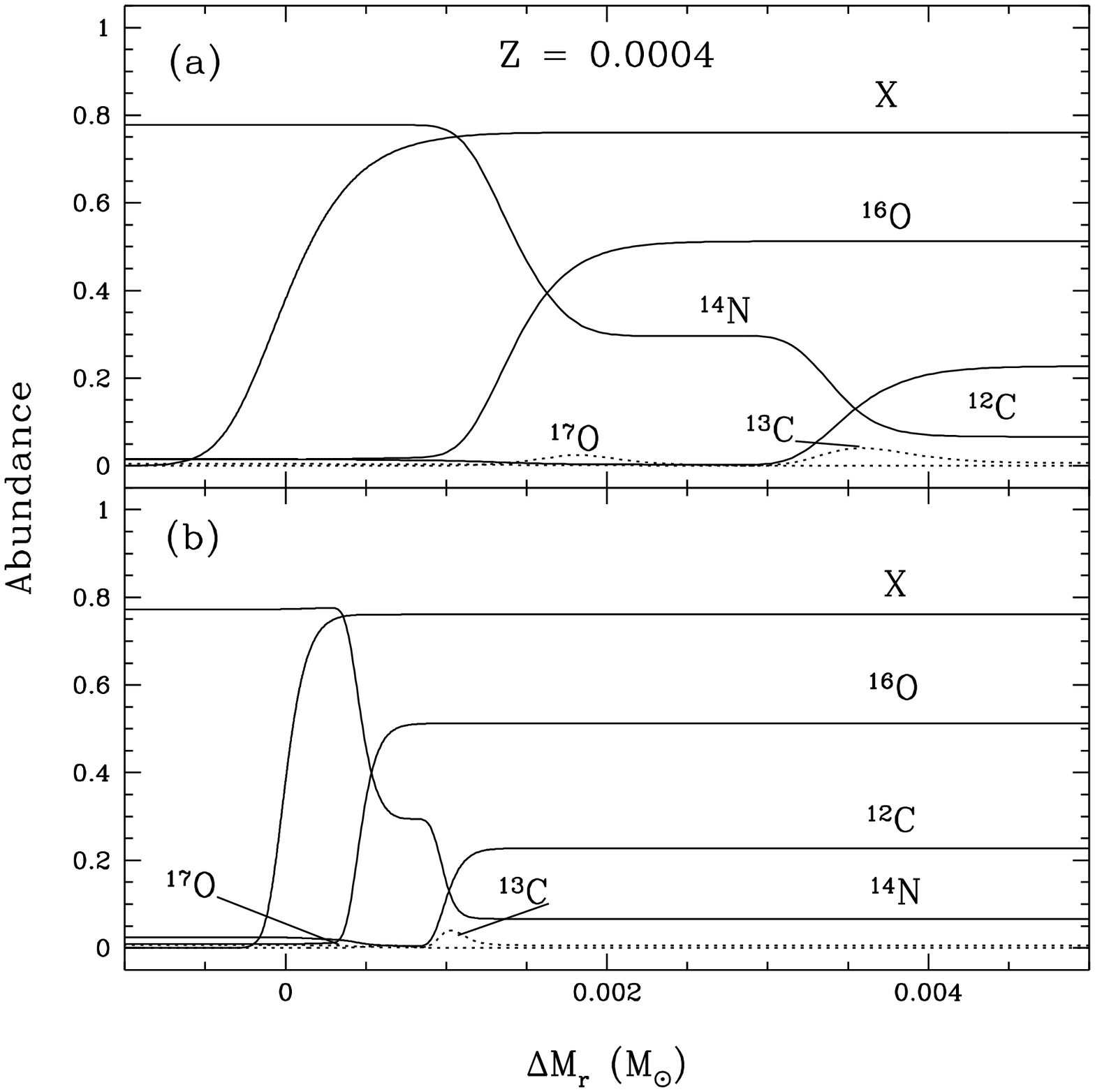}
\caption{As Figure 8, except for Z = 0.0004.}
\end{figure}
\begin{figure}
\plotone{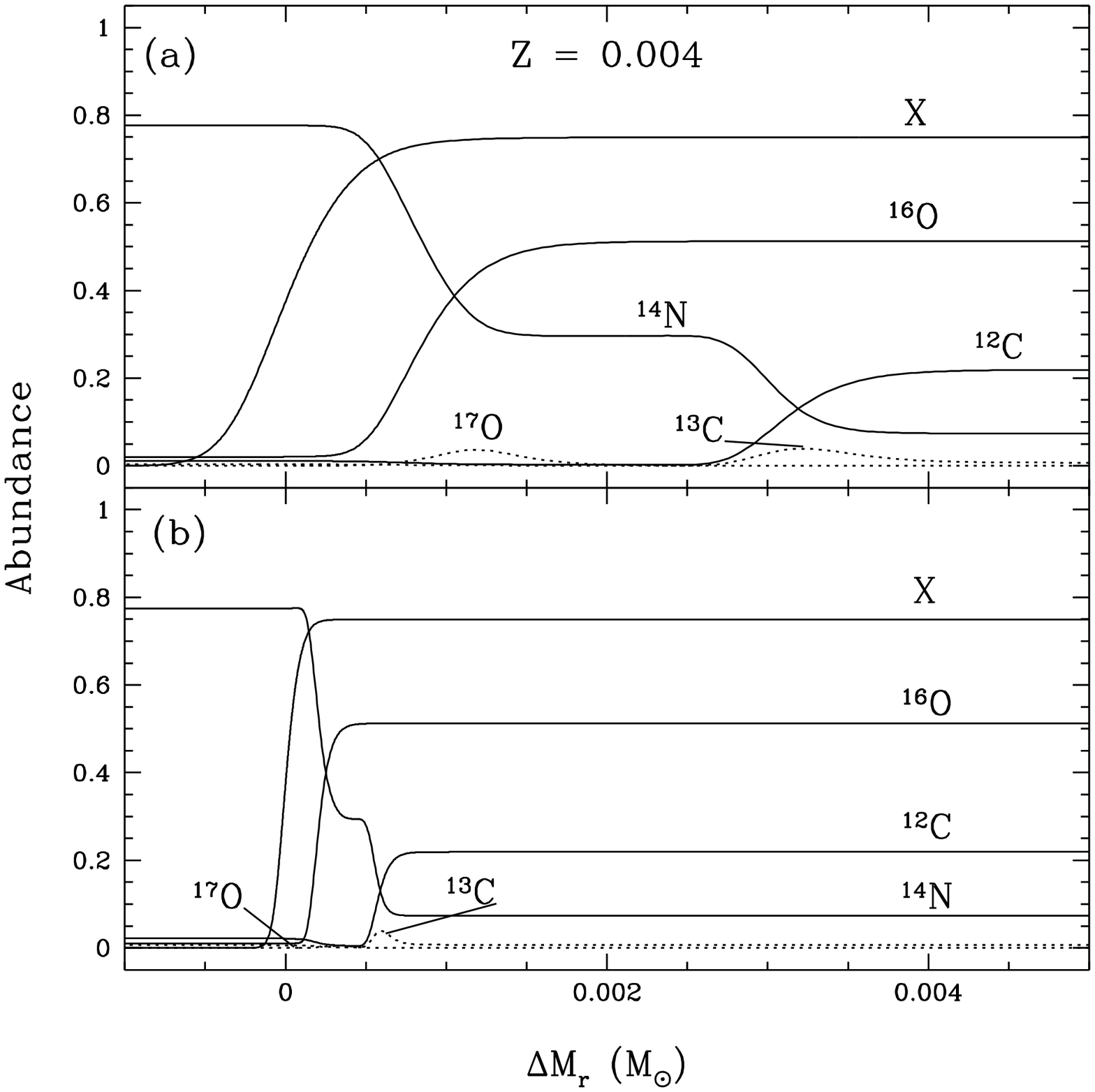}
\caption{As Figure 8, except for Z = 0.004.}
\end{figure}
\begin{figure}
\plotone{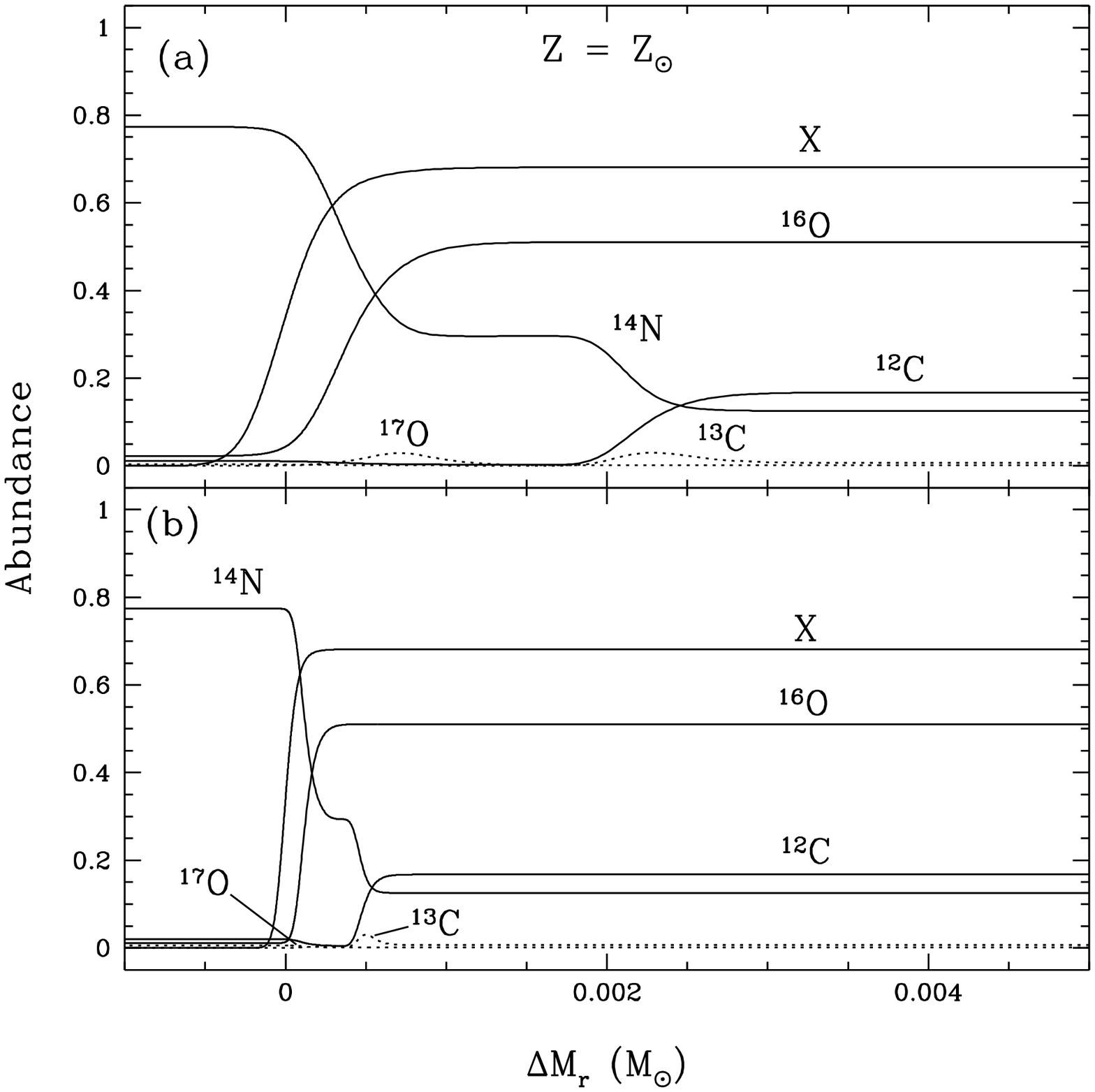}
\caption{As Figure 8, except for Z = Z$_\odot$.}
\end{figure}

The narrowing of the C- and O-depleted regions is compared in greater
 detail in Figure 12 for all the sequences at the points along the RGB
 where mixing is assumed to occur.
As the sequences evolve, the C- and O-depleted regions occupy less mass above
 the H shell.
This limits the amount of matter that can be processed and requires that the
 mixing penetrate closer to the H shell to reach the synthesized material.
In addition, we see that higher metallicities reduce the breadth of the 
 regions above the H shell from which synthesized matter can be mixed outward,
 as also found by SM79.
\begin{figure}
\plotone{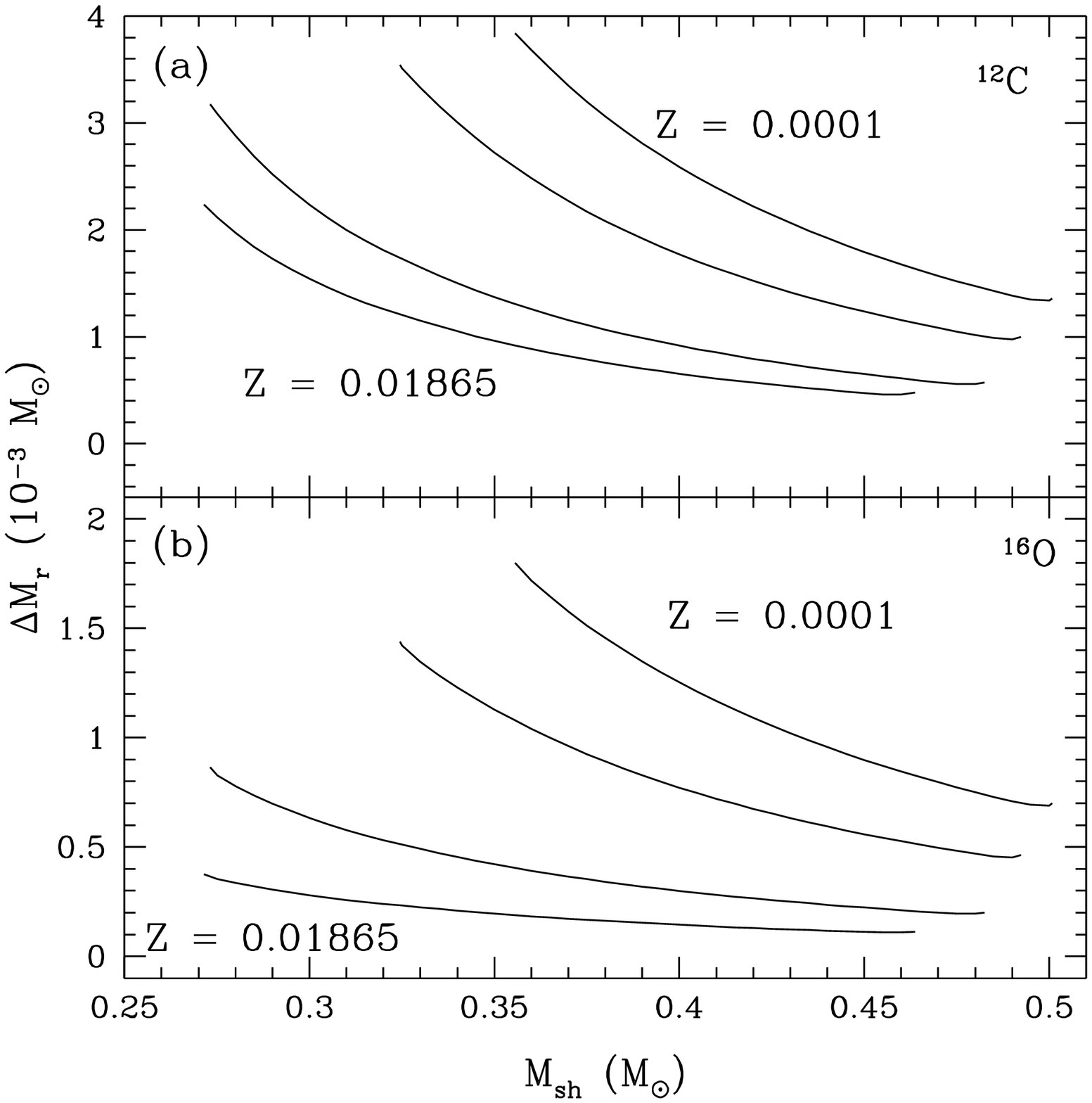}
\caption{The mass difference between the center of the
 H shell and the point where the (a) $^{12}$C and (b) $^{16}$O are equal to
 50\% of their envelope value during the evolution of the four RGB sequences.}
\end{figure}

From the previous section we know that the narrowing of the C- and O-depleted
 regions during the evolution up the RGB is accompanied by a steepening of 
 the H-shell profile.
This suggests that the substantial narrowing of these regions during the 
 RGB evolution might be less apparent if the CNO abundances are plotted
 in terms of the quantity ${\Delta}$S [given in equation (5)], since ${\Delta}$S
 measures mass difference in units of the H-shell thickness.
We illustrate this point in Figure 13, where we show the CNO abundance profiles 
 at both the start of mixing and the tip of the RGB for the Z = 0.0004 sequence.
We see that these CNO abundance profiles are nearly invariant during the RGB
 evolution when plotted as a function of ${\Delta}$S.
\begin{figure}
\plotone{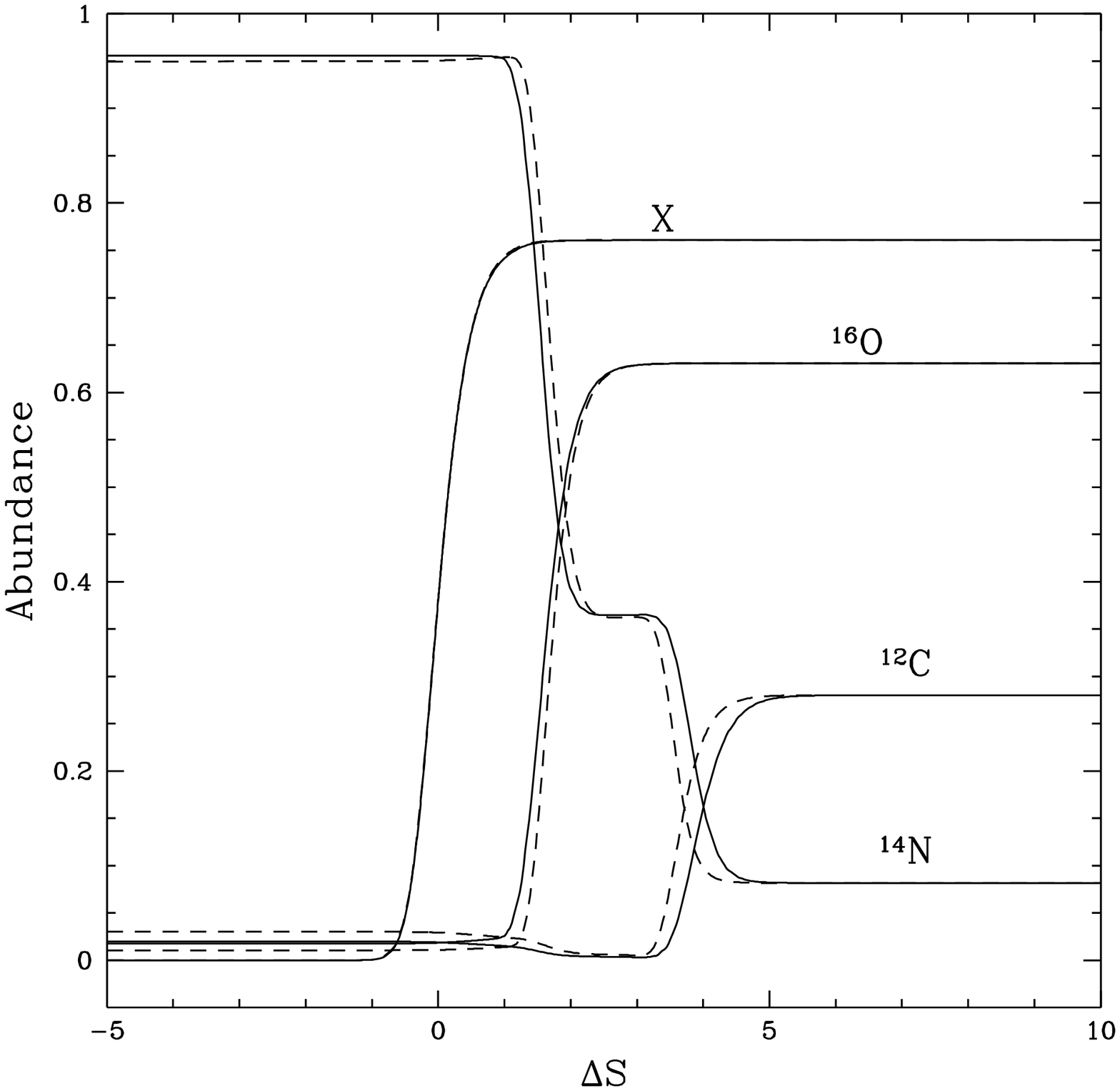}
\caption{Comparison of the H and CNO profiles for the Z
 = 0.0004 sequence as a function of ${\Delta}$S at the start of mixing (solid 
 line) and at the tip of the RGB (dashed line). The hydrogen-mass fraction is
 traced by the curved labeled X.}
\end{figure}
This invariance is useful when comparing and interpolating abundances between
 models of a single sequence.
In addition, ${\Delta}$S naturally lends itself to discussions of mixing depth.
For example, the ``top,'' ``center,'' and ``base'' of the H shell always
 appear at ${\Delta}$S = 1, 0, and $-$1, respectively, independent of any model.
A value of ${\Delta}$S = 0 corresponds exactly to a H depletion of 50\% of
 its envelope value, while ${\Delta}$S = 1 and $-$1 correspond to H depletions
 of approximately 2.5\% and 99.9\%, respectively.

The first reaction in the CN and ON cycles, before they attain equilibrium,
 creates $^{13}$C and $^{17}$O from $^{12}$C and $^{16}$O, respectively.
This causes the $^{12}$C/$^{13}$C ratio to dip below the equilibrium value near 
 4, before the other reactions in the CN cycle attain equilibrium with
 the $^{12}$C$({\it p,{\gamma}})^{13}$N$({\beta}^{+}{\nu})^{13}$C reactions,
 thereby building up $^{12}$C to its equilibrium value.
This behavior can be seen in Figures 8 - 11 and indicates that one would
 expect to see the $^{12}$C/$^{13}$C ratio decrease along the RGB beyond 
 the start of mixing provided that the mixing extends at least into the
 C-depleted region on an appropriate timescale.
In comparison, the $^{16}$O/$^{17}$O ratio is actually much below its 
 equilibrium value throughout most of the O-depleted zone as the $^{17}$O 
 abundance slowly enhances, then depletes near the H shell.
This is presented in Figure 14 where the $^{16}$O/$^{17}$O ratio is plotted
 for each sequence at the start of mixing and at the tip of the RGB.
\begin{figure}
\plotone{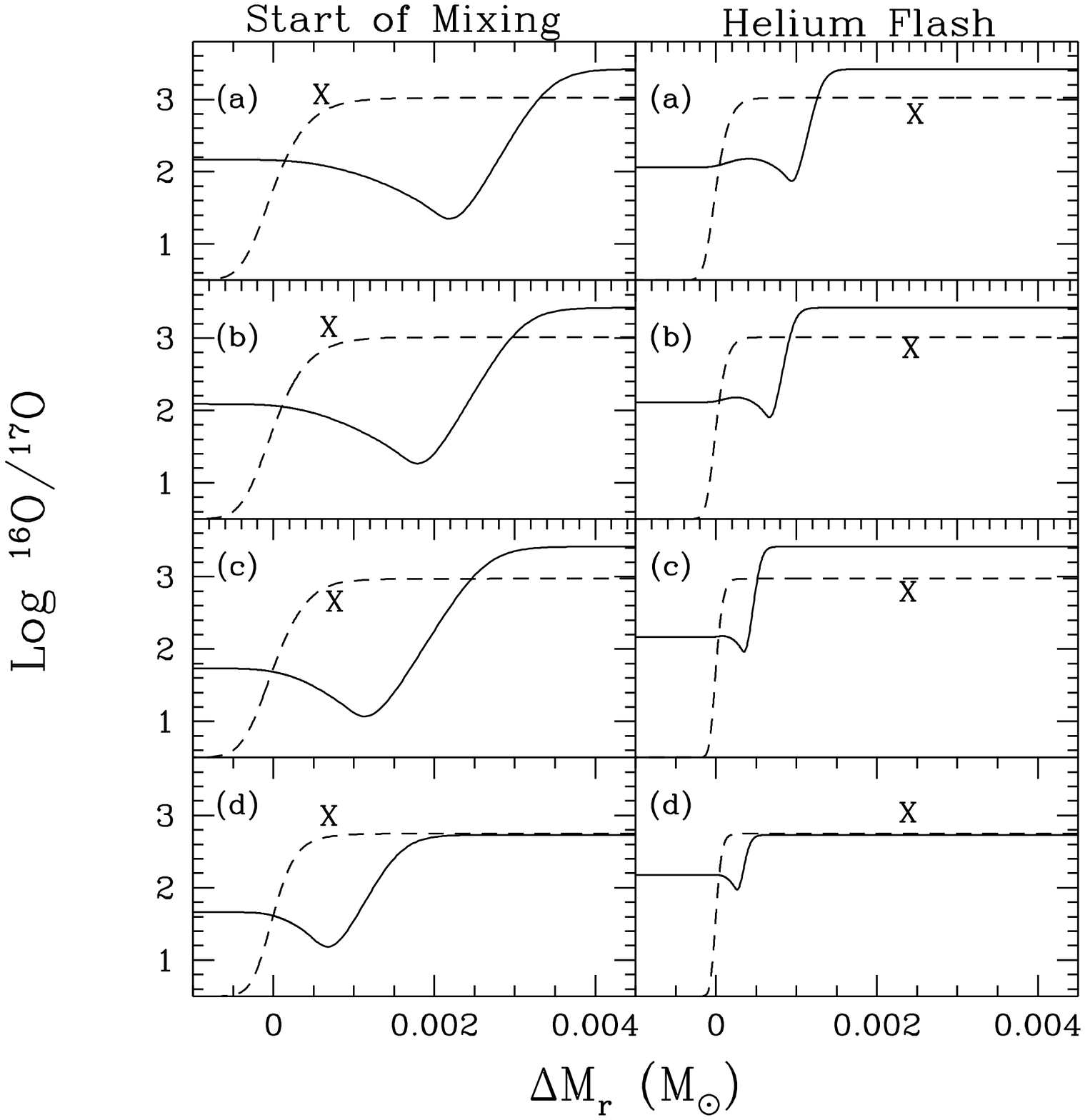}
\caption{The $^{16}$O/$^{17}$O ratio for the Z =  (a) 
 0.0001, (b) 0.0004, (c) 0.004, and (d) Z$_\odot$ sequences at the
 start of mixing (left panels) and the tip of the RGB (right panels).
 The dashed curve traces the H-mass fraction, X, on an arbitrary scale.}
\end{figure}
Due to its dependence on temperature, the $^{16}$O/$^{17}$O equilibrium 
 ratio varies depending on both the metallicity and luminosity of the model
 chosen.
This is different from the equilibrium ratio of $^{12}$C/$^{13}$C which is
 independent of temperature in the range relevant to these sequences.

\subsection{Na-Mg-Al}

We next examine the profiles of the NeNa- and MgAl-cycle isotopes, as
 shown in Figures 15 - 18.
Because most of the processing for these two cycles occurs close to or within
 the H shell, we plot these figures as a function of ${\Delta}$S for the
 sake of clarity.

The profiles for the Z = 0.0001 and Z = 0.004 sequences are slightly
 different from the ones presented in Paper I, since we have adjusted some of
 the relevant reaction rates, as discussed in section 6.
All of the sequences show the initial rise in $^{23}$Na from $^{22}$Ne well
 above the H shell at the start of mixing (panel a).
By the time the sequences have evolved to the helium flash (panel b), this
 initial Na plateau decreases in width as a result of the steepening 
 temperature gradient.
For the lowest metallicity sequence, the temperature is high enough at the
 onset of mixing to allow the NeNa cycle to begin building up $^{23}$Na
 from $^{20}$Ne at the top of the H shell (Figure 15a).
The NeNa cycle becomes important at the start of mixing for the Z = 0.0004
 sequence only inside the H shell (${\Delta}$S $\sim$ 0.6).
In the Z = 0.004 sequence, this extra production of Na occurs around 
 log(L/L$_{\odot}$) = 1.74 (M$_{\rm sh}$ = 0.275 M$_{\odot}$), soon after mixing
 begins; but, for the Z = Z$_\odot$ sequence, $^{20}$Ne never becomes
 a significant source of $^{23}$Na compared to its production from $^{22}$Ne.
Figures 15b - 17b show that the NeNa cycle feeds enough $^{20}$Ne into
 $^{23}$Na to substantially increase the Na production near the tip of the RGB
 for all Population II metallicities, but only in the lowest metallicity
 RGB tip model does this occur above the top of the H shell.
In the two lowest metallicity sequences, the NeNa cycle also experiences a
 leakage reaction through $^{23}$Na$({\it p,{\gamma}})^{24}$Mg causing an 
 enhancement in the $^{24}$Mg abundance.
As these two sequences evolve further up the RGB, the $^{23}$Na abundance within
 the H shell begins to diminish as leakage from the cycle depletes the
 $^{20}$Ne reserves.
Figure 15b demonstrates this behavior.
Approaching the H shell from above, the $^{20}$Ne/$^{23}$Na ratio first drops
 below the equilibrium ratio expected from the NeNa cycle, before the
 $^{23}$Na$({\it p,{\alpha}})^{20}$Ne reaction can build up the $^{20}$Ne
 abundance.
Around ${\Delta}S =$ 0.2, the leakage reaction begins to drain both the 
 $^{20}$Ne and $^{23}$Na, causing a local maximum at this point.
Since the NeNa cycling is less robust in the higher metallicity sequences, 
 this rise in the $^{20}$Ne abundance is less apparent.

The two lowest Z sequences show that $^{27}$Al enhancements at the onset of
 mixing come from the rapid depletions of $^{25}$Mg (by way of $^{26}$Al)
 and $^{26}$Mg inside the top of the H shell (Figures 15c and 16c).
On the contrary, the two highest Z sequences fail to produce any observable
 $^{27}$Al enhancements above the center of the H shell at the onset of mixing.
In fact, even at the tip of the RGB, the only Al that the Z = 0.004
 sequence can create comes from $^{25,26}$Mg.
For the Z = Z$_\odot$ sequence, $^{27}$Al is only modestly produced by
 a minor depletion of $^{26}$Mg at the tip of the RGB, while $^{25}$Mg is
 converted into $^{26}$Al at all luminosities.
In this sequence, ${\tau}_{\rm shell}$ is less than the decay time of $^{26}$Al
 (${\tau}$ = 7.4 x 10$^{5}$ yr), precluding any build-up of $^{26}$Mg and
 $^{27}$Al within or above the H shell.
As the metallicity decreases, however, the total $^{27}$Al increases 
 by substantially more than can be supplied by the initial $^{25,26}$Mg.
The source for this enrichment is $^{24}$Mg, which is itself replenished by
 leakage from the NeNa cycle.
By log(L/L$_{\odot}$) = 2.8 (M$_{\rm sh}$ = 0.425 M$_{\odot}$), the Z = 0.0001
 sequence begins to show the signs of enhancements in the $^{27}$Al abundance
 from $^{24}$Mg.
The Z = 0.0004 sequence begins its ``over-production'' of $^{27}$Al at a higher 
 luminosity on the RGB: log (L/L$_{\odot}$) = 3.2 (M$_{\rm sh}$ = 0.475 
 M$_{\odot}$).

The changes in the Al abundance are very sensitive to metallicity and
 luminosity (i.e. temperature) as shown in panels (c) and (d) of Figures 15 -
 18.
The reader should note that the ordinates of these panels have different
 scales for each sequence, with the lowest metallicity having the greatest
 range.
For example, Figure 16d shows that for the Z = 0.0004 sequence, the $^{27}$Al
 gained at the center of the H shell is +0.57 dex, which comes mostly
 from the decrease in $^{25,26}$Mg.
In comparison, the lowest metallicity sequence shows an increase in $^{27}$Al
 of +0.77 dex at the center of the H shell, of which only 0.42 dex comes from
 $^{25,26}$Mg.
The other 0.35 dex is supplied by $^{24}$Mg, either directly or through 
 leakage from the NeNa cycle.

\begin{figure}
\plotone{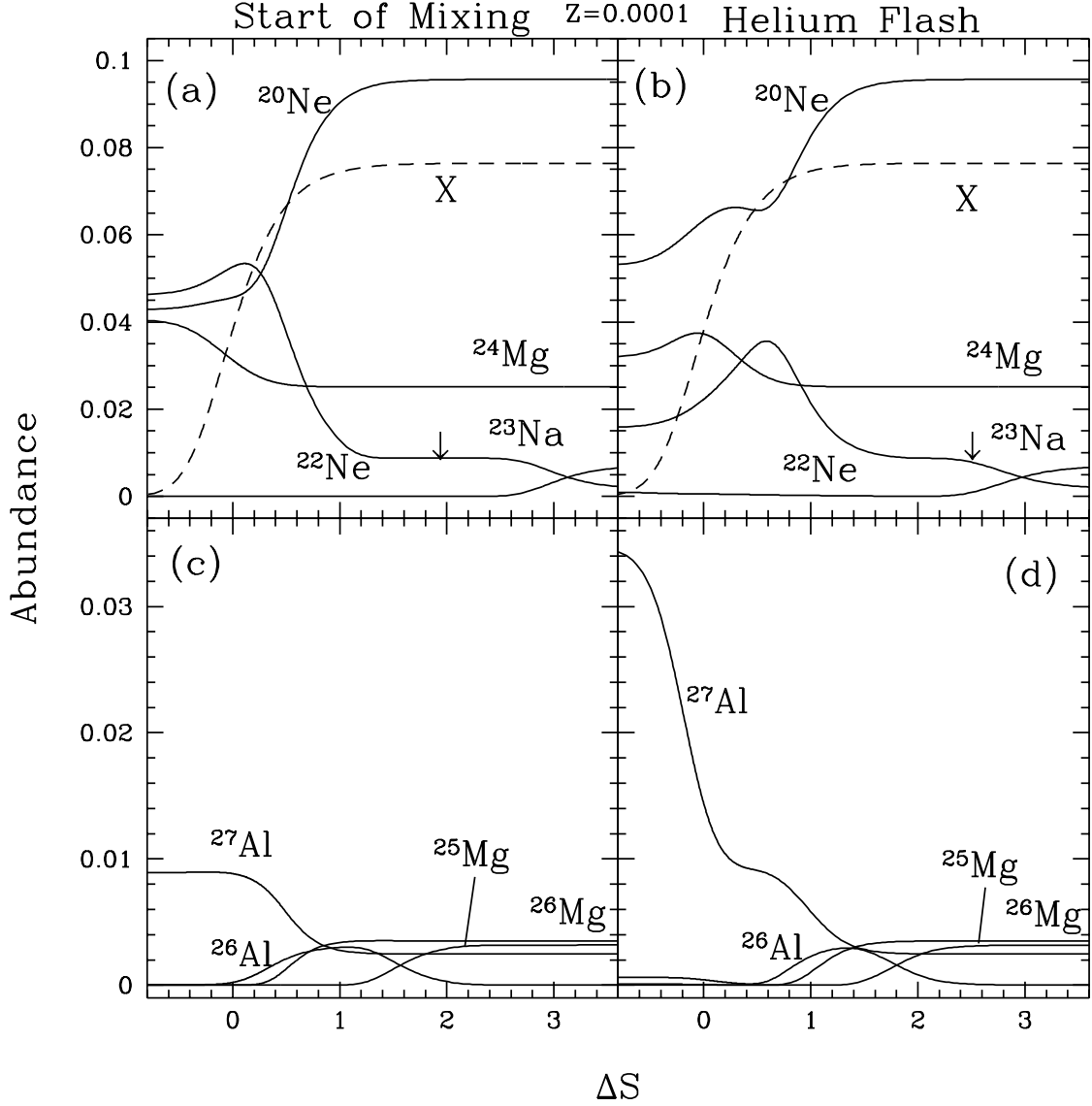}
\caption{The NeNa-cycle isotopes (top panels) and 
 MgAl-cycle isotopes (bottom panels) at the presumed start of mixing 
 (left panels) and at the tip of the RGB (right panels) for the
 Z = 0.0001 sequence.  The dashed line is the H-mass fraction scaled by
 a factor of 10.  The ordinate is the number abundance relative to all metals.
 The abscissa, ${\Delta}$S, is defined as the mass difference,
 ${\Delta}{\rm M_r}$, between any point and the center
 of the H shell divided by the H-shell thickness, ${\Delta}{\rm M_H}$.
 The arrows indicate the center of the O shell, defined
 as the point where the $^{16}$O abundance is equal to half of its envelope
 value.}
\end{figure}
\begin{figure}
\plotone{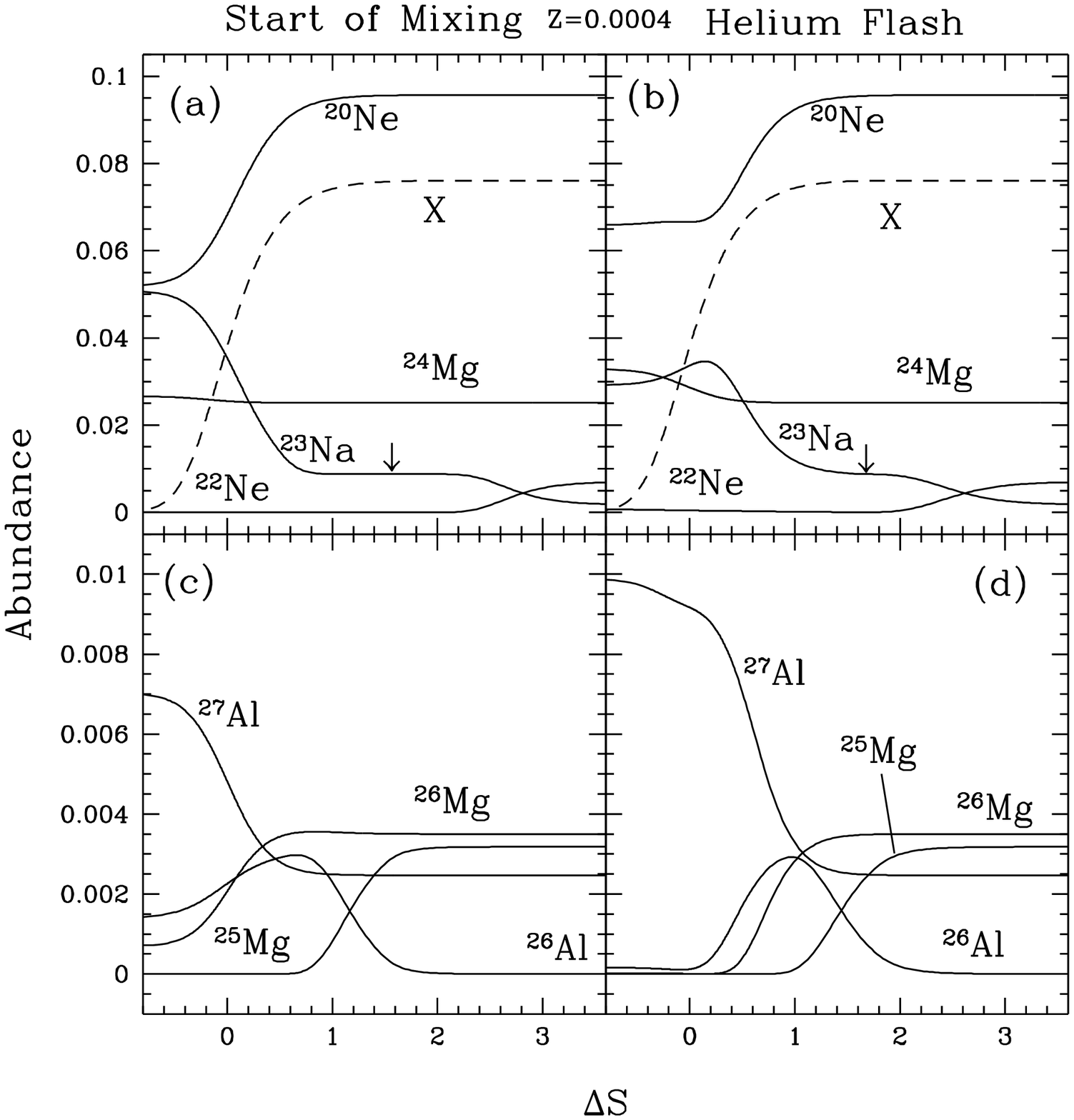}
\caption {As Figure 15, but for Z = 0.0004.}
\end{figure}
\begin{figure}
\plotone{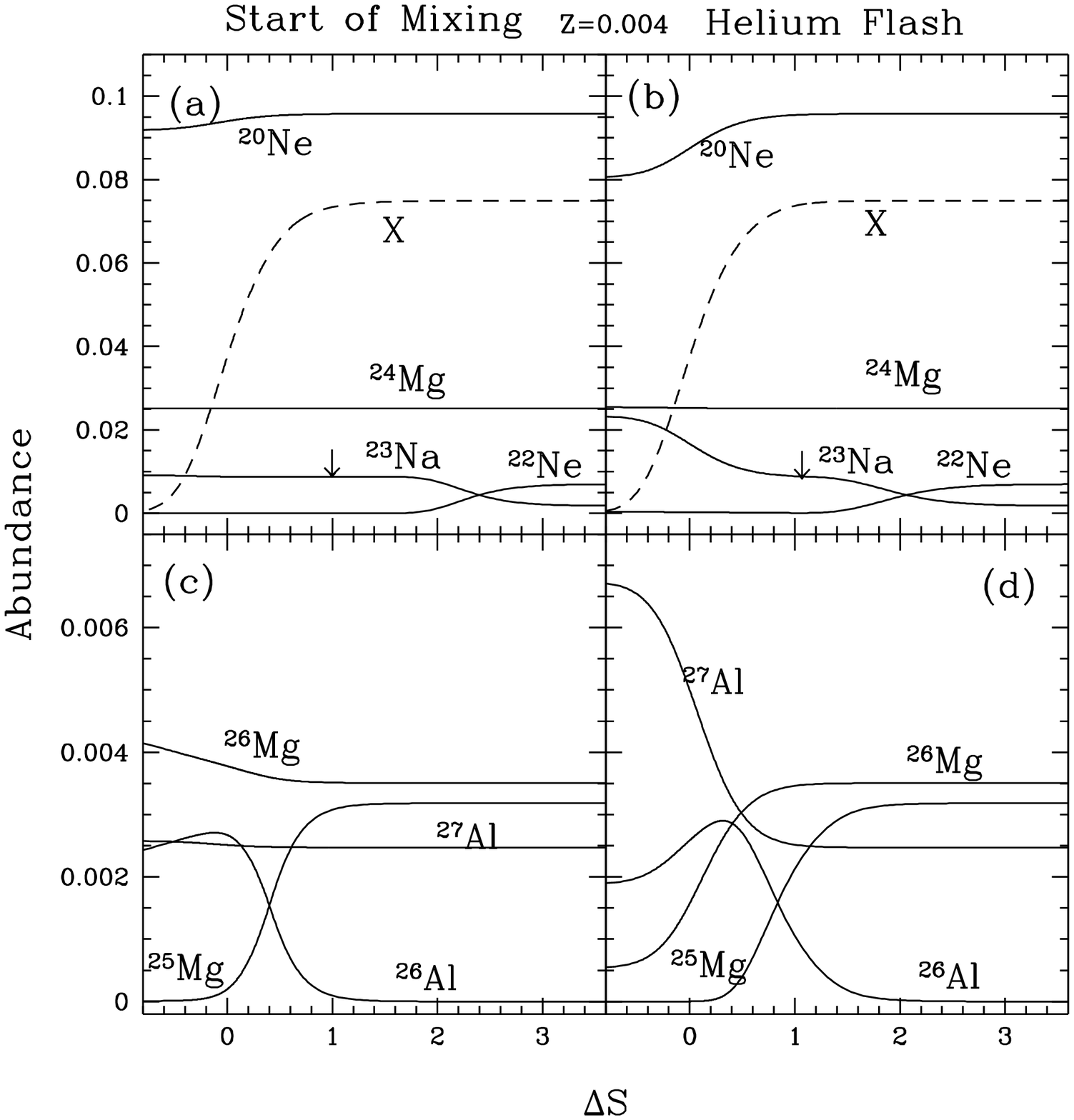}
\caption{As Figure 15, but for Z = 0.004.}
\end{figure}
\begin{figure}
\plotone{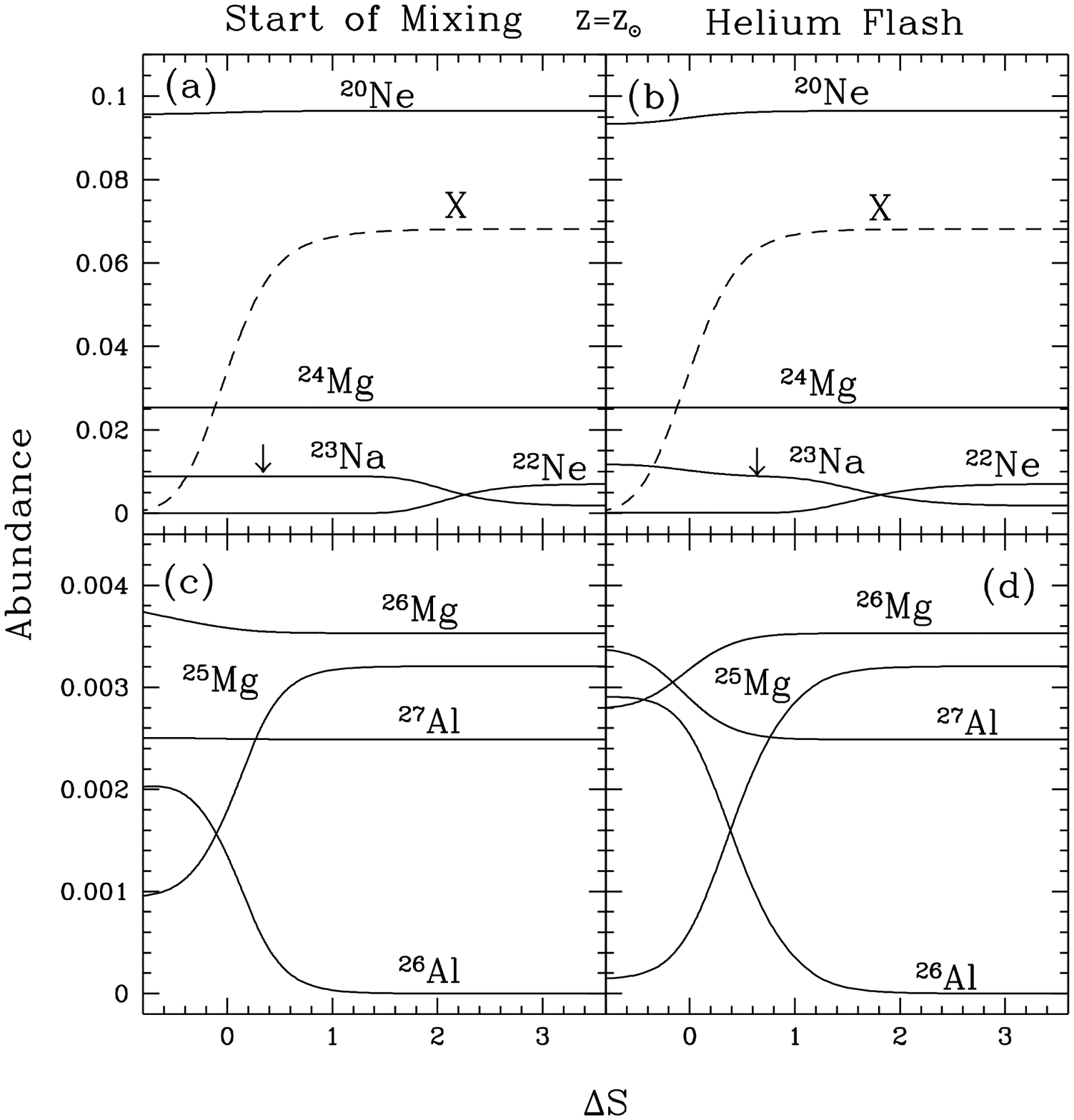}
\caption {As Figure 15, but for Z = Z$_\odot$.}
\end{figure}

\section{Consequences for Mixing}

We now look at the implications of these abundance profiles if the stars
 were to undergo mixing.
In order to fully understand the results, we must first address 
 the question of timescales.
There are several which affect the production and mixing of the elements.
The first is the shell burning timescale, ${\tau}_{\rm shell}$, which allows 
 the abundance profiles around the H shell to be set up.
This is determined by the rate at which the H shell moves outward in mass
 and is included explicitly in our calculations through the use of the
 stationary shell approximation.
The second is the characteristic timescale, ${\tau}_{\rm mix}$, over which the 
 mixing carries material away from the H shell to the convective envelope.
In the SM79 scenario, this is dependent upon the angular velocity and 
 angular momentum distribution in the RGB envelope.
Finally, there is the evolutionary timescale, ${\tau}_{\rm RGB}$, which
 determines for how long the mixing process can act and is defined by the time
 spent on the RGB after the H shell has burned through the H discontinuity.
SM79 pointed out that the mixing process is gradual in that the timescale for
 changing the envelope composition does not need to be less than the 
 shell burning timescale.
If it were this rapid, there would be abrupt changes in the surface composition 
 shortly after the onset of mixing.
This is not supported by the observations.
Rather, the relevant criterion is that the amount of nuclearly processed
 matter mixed outward from the shell over the RGB evolutionary timescale be
 equal to the envelope mass; i.e., that most of the envelope passes
 through the nuclear burning regions by the time the star reaches the tip of 
 the RGB.
This is what causes the gradual changes in the abundances with increasing
 luminosity (see, e.g., Bell, Dickens, \& Gustafsson \markcite{r3}1979 for
 M92 and NGC 6397; Carbon et al. \markcite{r4} 1982 for M92; Trefzger et al.
 \markcite{r7}1983 for M15; Briley et al. \markcite{r6}1990 for NGC 6397;
 Briley et al. \markcite{r16}1997 for M71).

If mixing were to occur on an appropriate timescale, what changes would we
 expect to see based on these sequences?
The answer, of course, depends on the depth of mixing.
For example, all of the sequences behave similarly regarding the CNO abundances.
The distinguishing characteristic which separates one sequence from another
 is the distance above the H shell of the C- and O-depletion regions as
 discussed in section 4.1, and shown in Figures 8 - 11.
Thus, we anticipate that there would be C vs. N and O vs. N 
 anticorrelations at all metallicities, with the extent of their variations
 depending on the depth of mixing.
Along with these anticorrelations, we would expect to see the $^{12}$C/$^{13}$C
 ratio near its equilibrium value of 4 if mixing penetrated far enough
 into the C-depleted region to cause a C vs. N anticorrelation.
Likewise, if mixing reached deeply enough to cause an O vs. N anticorrelation,
 we would expect to see the $^{16}$O/$^{17}$O ratio near its equilibrium value
 between $\sim$ 50 and 150, depending on the metallicity and luminosity of the
 model.
The observations show C vs. N anticorrelations and $^{12}$C/$^{13}$C ratios
 near equilibrium in clusters of all metallicities (see, e.g., Norris
 \markcite{r21}1981; Smith \& Norris \markcite{r23}1983; Smith \& Bell
 \markcite{r26}1986; Pilachowski \markcite{r64}1988; Briley et al.
 \markcite{r15}1989; Smith \& Mateo \markcite{r27}1990; Sneden et al.
 \markcite{r65}1991; Smith et al. \markcite{r69}1996; Smith et al.
 \markcite{r128}1997), but there is less
 evidence of O vs. N anticorrelations in the more metal-rich clusters 
 (Dickens et al. \markcite{r78}1991; Sneden et al. \markcite{r71}1994;
 Briley, Smith, \& Lambert \markcite{r47}1994; Norris \& Da Costa
 \markcite{r46}1995a; Briley et al. \markcite{r122}1997).
There is yet no data available for the $^{16}$O/$^{17}$O ratio in globular
 cluster stars.
This constrains both ${\tau}_{\rm mix}$ and the depth of mixing for the
 higher metallicity clusters.
Perhaps the trends seen in the higher metallicity clusters are the result
 of narrower shell processing regions which allow less material to
 be synthesized in a given time.
In any case, the metallicity at which the results of the O-N conversion become
 commonplace seems to be between the Z = 0.0004 sequence (for example, M13
 [KSLS93]) and the Z = 0.004 sequence (M71 [Sneden et al. \markcite{r71}1994; 
 Shetrone \markcite{r114}1996a]).

Understanding the implications of the Na, Mg, and Al isotopic abundances
 is complicated by the fact that their yields are very sensitive
 to the sequence metallicity and model luminosity. 
In fact, the only similarity among the sequences is the production of
 $^{23}$Na from $^{22}$Ne in a plateau region above the H shell.
This region becomes more narrow and moves closer to the H shell with both
 increasing luminosity and metallicity.
Figures 15 - 18 show the initial change in the Na abundance relative to the
 center of the O shell, which we define as the point where the $^{16}$O
  abundance is equal to half of its envelope value, and mark with an arrow
  above the $^{23}$Na curve.
The reader may notice that when shown in ${\Delta}$S-space, the center of
 the O shell actually moves ${\em away}$ from the center of the H shell
 with increasing luminosity, indicating that the H shell is steepening more
 quickly than the O-depleted region is narrowing.
At lower luminosities the Na is enhanced above the center of the O shell.
This leads to a scenario where Na can be increased without altering the O
 abundance as observed in the metal-poor stars of $\omega$ Cen by
 Norris \& Da Costa (\markcite{r46}1995a) and in the metal-rich cluster NGC
 362 by Briley et al. (\markcite{r122}1997).
As the sequences evolve, the Na-enriched region overlaps the O-depleted
 region and helps to describe the global anticorrelation between O and Na
 (Norris \& Da Costa \markcite{r86}1995b; Kraft et al. \markcite{r113}1997).
The $^{23}$Na abundance becomes more strongly enhanced as the result of the NeNa
 cycle only in the two most metal-poor sequences and only within or very close
 to the H shell.
To reach this abundant supply of Na would require mixing deeply into the
 O-depleted region, implying that large Na enhancements should be accompanied by
 large O depletions, as seen in M13 (Kraft et al. \markcite{r113}1997)

For the other isotopes in the NeNa and MgAl cycles, it is instructive to
 look at a comparison of two sequences at a common luminosity in order to
 gain some insight into how the metallicity affects the interpretation of
 the models.
In Figure 19, we plot the NeNa (panel a) and MgAl (panel b) isotopes for the
 two metallicities which represent the extremes in our sequences.
\begin{figure}
\plotone{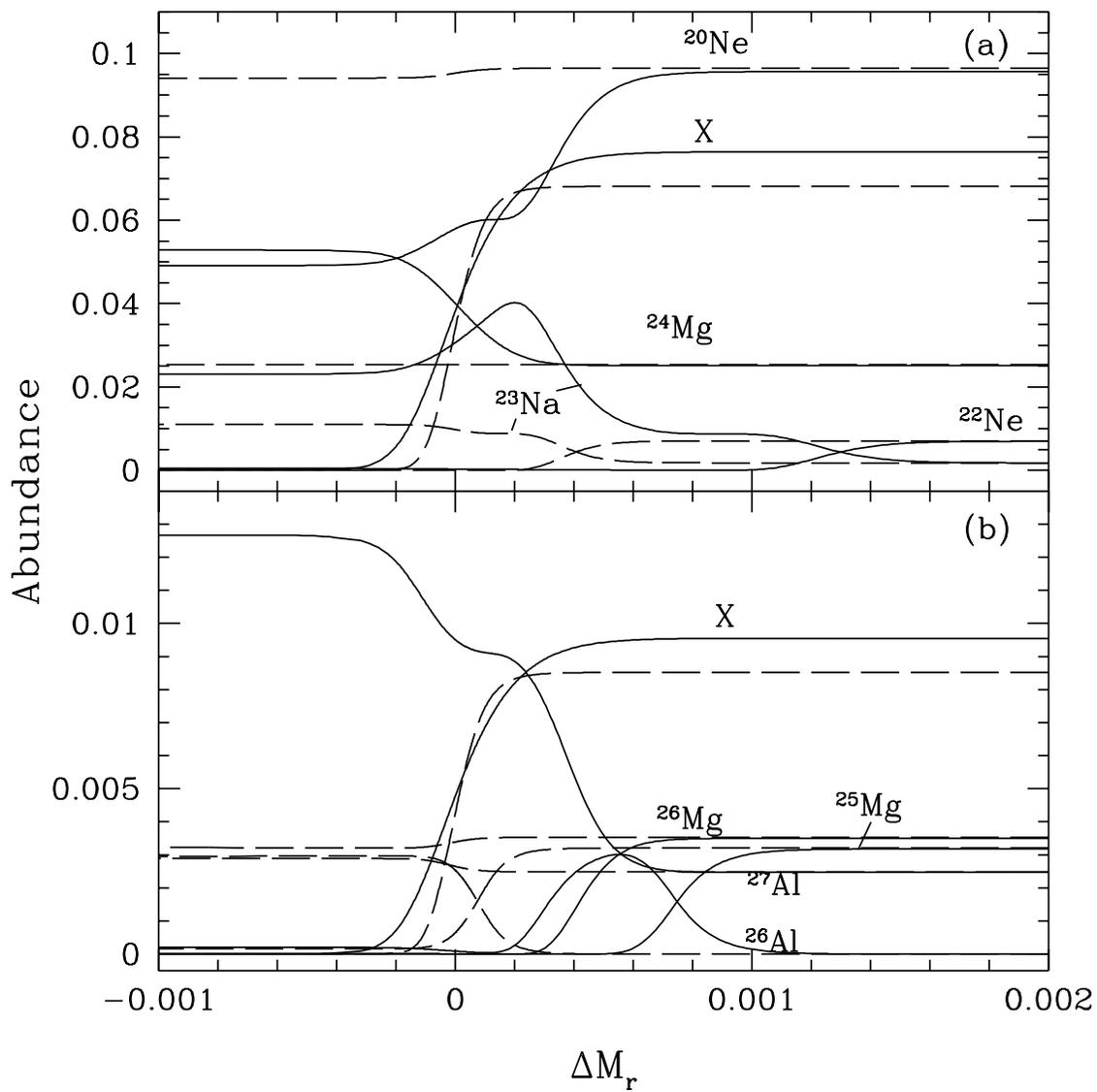}
\caption{(a) The NeNa cycle isotopes for the Z = 0.0001
 (solid line) and Z = Z$_\odot$ (dashed line) sequences at log(L/L$_\odot$) =
 3.0.  (b) The MgAl cycle abundances for the same models in
 panel (a). The hydrogen-mass fraction, X, is scaled by a factor of 10 in 
 panel (a), and by a factor of 80 in panel (b).}
\end{figure}
Clearly there is much more processing occurring in the low metallicity sequence
 where the temperatures are higher.
The high metallicity sequence shows less robust processing: $^{23}$Na is 
 slightly enhanced from $^{22}$Ne and Al is increased mostly in the form of 
 the meta-stable isotope of $^{26}$Al which decays to $^{26}$Mg.
Thus, deep mixing in the metal-richest sequence would produce little changes in 
 these elements unless the $^{26}$Al could be mixed to the surface in large
 quantities before decaying.
If ${\tau}_{\rm mix}$ is longer than the $^{26}$Al half-life, then the 
 $^{26}$Mg/$^{25}$Mg ratio would be seen to increase in a metal-rich, deep
 mixing scenario.
As the metallicity decreases, the $^{27}$Al does begin to build up from both
 $^{25,26}$Mg (see, e.g., Figure 17d) due to the increasing temperatures.
The metal-poor sequences are able to produce more $^{27}$Al than can
 be accounted for by $^{25,26}$Mg, as demonstrated by Figure 19b.
The extra Al enhancements occur on the upper RGB at the base of the H shell.
Therefore, in order for the depletion of Mg and the large Al enhancements to
 be seen at the stellar surface, mixing must occur deep within the H shell
 during the latter half of the RGB lifetime.
The corollary to this is that if large Al enhancements are seen, they should
 be accompanied by large increases in N and Na, and large decreases in
 $^{25,26}$Mg and $^{16}$O.
(We discuss the effect on $^{24}$Mg later.)
Since the changes occur only on the upper RGB, there are now tighter constraints
 on the mixing time.

Our results are qualitatively consistent with the relatively few observations
 of total Mg abundances.
That is, deep in the H shell where there are large Al enhancements,
 the total Mg abundance is depleted, in accord with the observations of 
 M13 RGB tip giants as seen by Shetrone (\markcite{r116}1996b), PSKL96, and
 Kraft et al. (\markcite{r113}1997).
This is in spite of the leakage from the NeNa cycle which creates a surplus
 of $^{24}$Mg.
For example, at the tip of the RGB in the Z = 0.0004 sequence, the total
 Mg abundance at the center of the H shell decreases by 9.9\% relative to
 the envelope value, while the $^{24}$Mg abundance increases by 14\%.
Only if mixing were to penetrate more deeply than this would the total Mg
 abundance become enhanced.

On the other hand, our models are at odds with the observations of the Mg
 isotopes in M13 and NGC 6752 by Shetrone (\markcite{r116}1996b,
 \markcite{r124}1997).
Whereas he observes a decrease in $^{24}$Mg which anticorrelates with
 Al, we predict an increase in $^{24}$Mg as previously noted.
This increase, of course, is dependent on the accuracy of the
 $^{23}$Na$({\it p,{\gamma}})^{24}$Mg rate as discussed in section 6.2.
In addition, the Shetrone results show a low abundance of both $^{25,26}$Mg
 when Al is low, and an increased abundance when Al is enhanced.
This is difficult to explain given our present understanding of the
 nuclear reaction rates.
We show that when Al is slightly enhanced, as in the star L598 in M13,
 the isotopes which create the Al are $^{25,26}$Mg.
However, when Al is substantially increased (${\Delta}$Al $\sim +1.0$ dex),
 we predict that $^{24}$Mg is the source of that enhancements {\em and} that
 both $^{25,26}$Mg remain depleted.
The Shetrone observations show that $^{24}$Mg is indeed the source
 of the large Al enhancements, but they also show that the other two
 isotopes are somehow increased.
Since,  the $^{24}$Mg proton-capture rate is much slower than either
 the $^{25}$Mg or $^{26}$Mg rates at the canonical temperatures in the H shell,
 any $^{24}$Mg depletions should be accompanied by a permanent diminution of
 $^{25,26}$Mg.

$^{25,26}$Mg become important at lower temperatures.
$^{25}$Mg is easily converted into $^{26}$Al in all sequences.
Only in the lower luminosity Population I models is it not completely exhausted.
Decreasing the metallicity of a sequence extends ${\tau}_{\rm shell}$
 and allows the resultant $^{26}$Al to decay into $^{26}$Mg before the shell
 overtakes the abundance profiles, as in the higher metallicity sequences.
$^{26}$Mg, on the other hand, requires higher temperatures before it can
 be processed into $^{27}$Al.
These last two points explain why $^{26}$Mg is seen to increase below the H
 shell in the solar-metallicity sequence.
The implications for the two Mg isotopes is that one should expect to see
 an anticorrelation between $^{25,26}$Mg and $^{26,27}$Al which becomes more
 pronounced at lower metallicities and higher luminosities.
Whether this can be seen in the higher metallicity sequences is also
 a matter of how much mixing can occur in these stars and is the subject
 of future work.

Finally, we look at the effect of all three Mg isotopes in the context
 of recent observations, which might be explained in the following manner.
For stars on the lower RGB or with high metallicity where the temperatures
 are too low to process $^{24}$Mg and $^{23}$Na, the Al abundance is 
 dominated by the destruction of $^{25,26}$Mg.
In these stars we expect to see an anticorrelation between the total
 Mg and Al abundances.
Because the heavier two Mg isotopes are depleted in the region of large
 Na production and O depletion, this would produce an anticorrelation
 between Na and Mg and a correlation between O and Mg as well, as seen
 by PSKL96.
However, since the Na is initially increased in the $^{22}$Ne-burning 
 plateau above the region of Mg depletion, it should not be surprising to
 see some increase in Na before seeing a change in Mg.
Whether or not the spread in the Mg abundance is correlated with the small
 increases in Na depends on the depth of mixing.
For metal-poor RGB tip stars, where the NeNa cycle begins leaking strongly
 into the MgAl cycle, we expect both the Al and Mg abundances to be controlled
 by $^{24}$Mg after the initial $^{25,26}$Mg has been depleted.
According to the nuclear reaction rates we have used, these stars would
 exhibit a $^{24}$Mg-Al correlation and either a Mg-Al correlation or
 anticorrelation, depending on the depth of mixing.
We discuss possible remedies for the discrepancies between our results and
 the observations in section 6.

One test of the significance of leakage into the MgAl cycle can be achieved
 by checking the constancy of [(Mg+Al)/M] among the cluster stars\footnote{In
 this instance, we use M to refer to the number abundance of metals.}. 
Any leakage would cause this quantity to show star-to-star variations.
Shetrone (\markcite{r114}1996a) has shown that the stars in M13 conserve the
 total number of Mg and Al nuclei.
However, this hypothesis is difficult to test since, as we show in Figure 20, 
 the total variation in the number of MgAl cycle nuclei due to leakage is
 only 0.045 dex at the center of the H shell at the tip of the sequence which
 represents the metallicity of M13, and is much less at lower luminosities.
\begin{figure}
\plotone{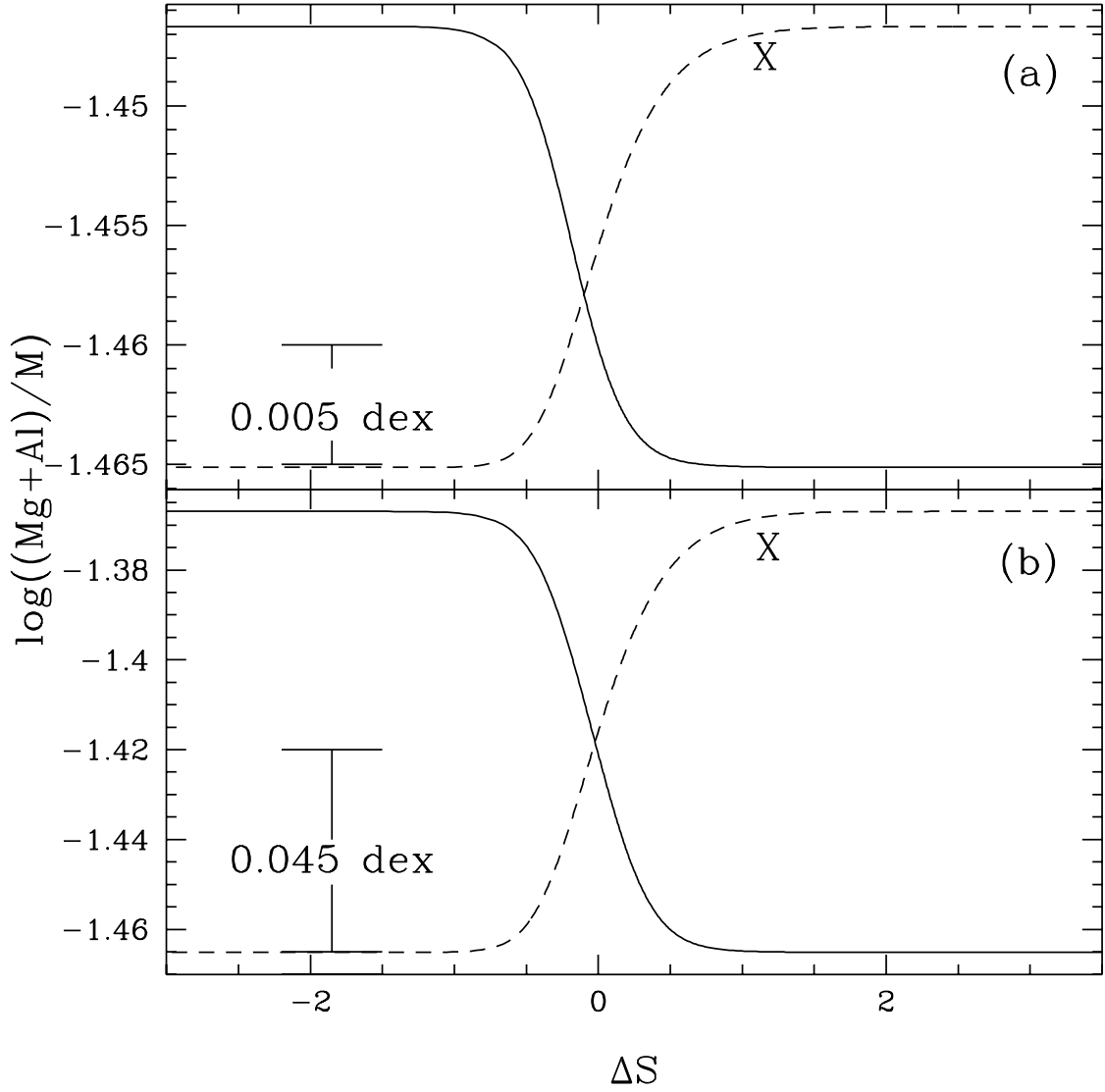}
\caption{The variation in the total Mg+Al nuclei due to
 leakage from the NeNa cycle for the Z = 0.0004 sequences at (a) the onset
 of mixing (b) the tip of the RGB.}
\end{figure}
A variation of this magnitude is typically much less than the uncertainties
 in the abundance measurements, and therefore, not likely to be observed.

\section{Nuclear Reaction Rate Uncertainties}

Our nuclear burning code includes some changes to reflect modern results for
 the nuclear reactions involved in the CNO, NeNa, and MgAl cycles.
In this section we explore the uncertainties in the reactions used in the
 code and discuss their effect on our results.

\subsection{$^{17}$O Proton Captures}

In the ON cycle, we substituted updated rates for the
 $^{17}$O$({\it p},{\alpha})^{14}$N and $^{17}$O$({\it p},{\gamma})^{18}$F
 reactions based on the formulae from Landr\'{e} et al. (\markcite{\r105}1990).
These rates are based on indirect measurements of the proton width of the 
 resonant state at 66 keV.
The formulae include two factors, ${\it f_1}$ and ${\it f_2}$, that 
 reflect the range of error in the measurement of the resonance at 66 keV
 and a higher level resonance in $^{18}$F, respectively.
For each decay channel ${\it f_1}$ is given a value between 0.2 and 1, and
 ${\it f_2}$ a value between 0 and 1.
Blackmon et al. (\markcite{r106}1995) have since directly measured the
 cross section at energies near the 66 keV resonance of 
 $^{17}$O(${\it p},{\alpha})^{14}$N and concluded that the total reaction 
 proceeds at approximately 10 times the rate given by CF88.
Therefore, we used the formulae given by Landr\'{e} et al. 
 (\markcite{r105}1990), but we extended ${\it f_1}$ down to a value of 0.05
 to allow the alpha-decay reaction rate to fall in line with the direct
 measurements of Blackmon et al. (\markcite{r106}1995).
Since the resonance at 66 keV dominates both rates for
 $0.01 < {\rm T_9} <$ 0.07, the value of ${\it f_2}$ is inconsequential;
 as such, we adopted ${\it f_2} =$ 0.1.
Figure 21 compares the cross-sections of Landr\'{e} et al.
 (\markcite{r105}1990), using two values of ${\it f_1}$, with the rates of CF88.
The dashed lines represent the value of 0.3 chosen in Paper I, while the
 dotted lines show the cross-sections used in this work.
It is clear from panel (a) that our rate for the
 $^{17}$O$({\it p},\alpha)^{14}$N reaction is around ten times that of CF88.
Since there are no new data for the $^{17}$O$({\it p},\gamma)^{18}$F reaction
 rate, we use the same values for
 ${\it f_1}$ and ${\it f_2}$ as in the $\alpha$-decay rate and determine that
 this new rate agrees fairly well with that of CF88, as shown in Figure 21b.
The most significant impact of the former rate is on the $^{16}$O/$^{17}$O 
 equilibrium ratio.  
Using the presently modified rates of Landr\'{e} et al. (\markcite{r105}1990)
 yields a temperature-dependent ratio between 50 and 150 in the range
 $0.03 \lesssim T_9 \lesssim 0.06$ with a peak near $T_9$ = 0.045.
This is at least a factor of ten larger than predicted with the rates of
 CF88.  

\begin{figure}
\plotone{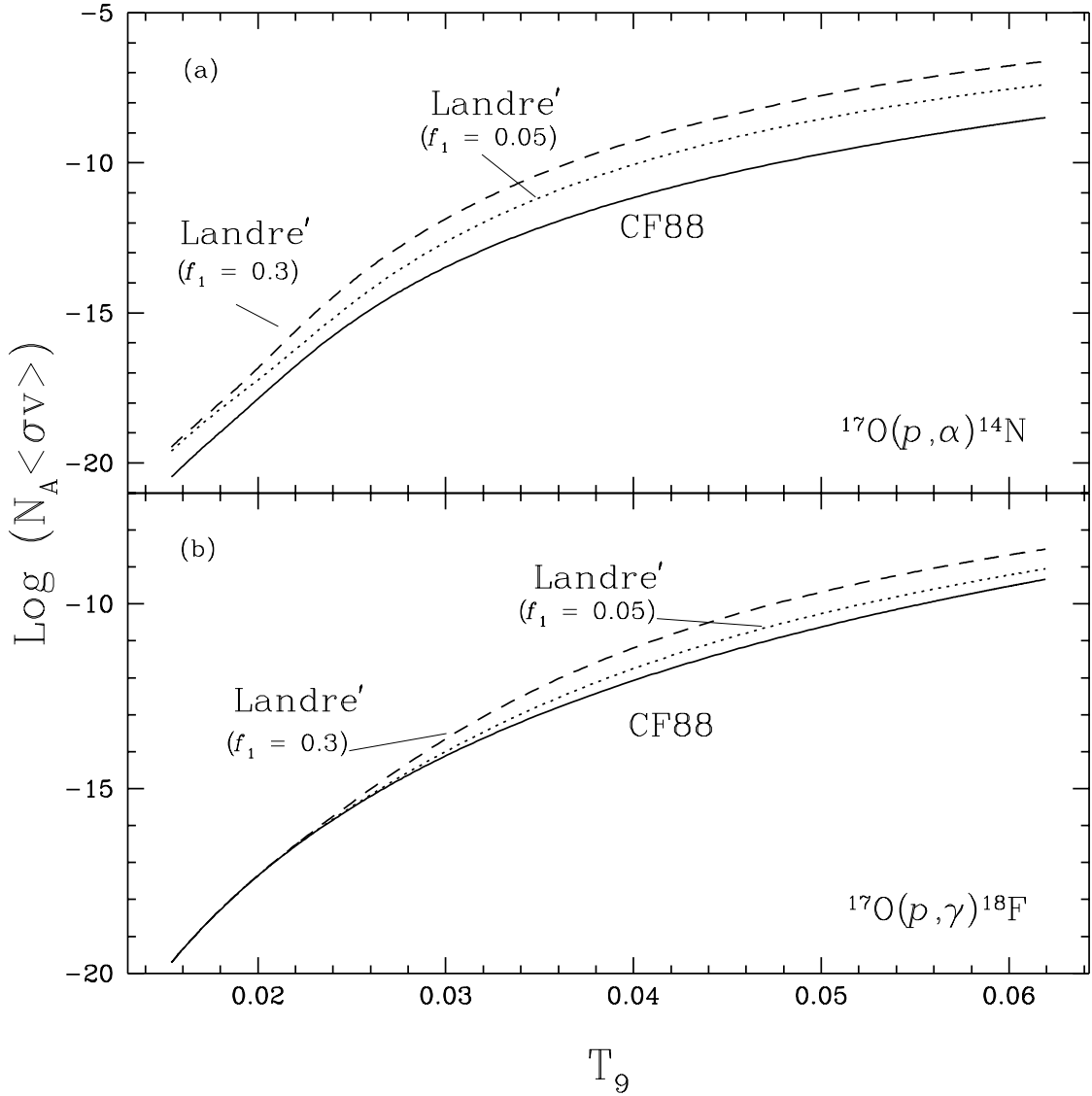}
\caption{A comparison of the new O proton-capture rates
 from Landr\'{e} et al. (1990) with those of CF88 for two values
 of ${\it f_1}$.  The ordinate is in logarithmic units of ${\rm cm^{3}\  s^{-1}
 \ mole^{-1}}$.}
\end{figure}

\subsection{NeNa Cycle Reactions}

We have replaced the rates of CF88 in the NeNa cycle by those of
 El Eid \& Champagne (\markcite{r107}1995; hereafter, EC95).
As done by CF88, EC95 modified some terms in their rates by a factor which
 ranges from 0 to 1 to reflect the uncertainty in some resonances.
Following the lead of CF88, we use an intermediate value of 0.1 for all such
 terms and check the effect of doing so on our analysis.
The rate of the first reaction in the cycle,
 $^{20}$Ne$({\it p},{\gamma})^{21}$Na, is very well determined.
This reaction is followed by a ${\beta}^{+}$-decay, which is very rapid 
 compared to the reactions in the cycle, as are most of the decays in both the
 NeNa and MgAl cycles.
The next reaction in the cycle is $^{21}$Ne$({\it p},{\gamma})^{22}$Na.
Since there is no appreciable source of $^{21}$Ne before the cycle starts and
 since this reaction is faster than the $^{20}$Ne proton capture by orders
 of magnitude at the H-shell temperatures, any uncertainty in its rate is
 inconsequential to the production of $^{23}$Na.
On the contrary, while $^{22}$Ne$({\it p},{\gamma})^{23}$Na proceeds much
 more rapidly than the $^{20}$Ne proton capture, there is an appreciable source
 of $^{22}$Ne in the envelope; therefore, it is important to understand this
 rate well, particularly in the higher metallicity sequences where the
 Na production might be dominated by $^{22}$Ne.
As discussed in EC95, the increase of the lowest resonance by 6 keV to 36 keV
 causes the lower limit for this rate to be 5-30 times faster than the
 CF88 rate in the range $T_9$ = 0.02-0.05.
This is the reason why the Na plateau exists so much further above the center
 of the O shell than originally suggested by the CF88 rates: the EC95 $^{22}$Ne 
 proton-capture rate proceeds nearly a factor of 40 faster than 
 $^{16}$O$({\it p,{\gamma}})^{17}$F at $T_9$ = 0.035, as opposed to nearly a
 factor of unity with the CF88 rates.
In addition, two resonances near 68 keV and 100 keV can cause this rate
 to be as much as a factor of two higher at $T_9$ = 0.03 than the suggested
 lower limit that we have used in the code.
Using the upper limits to this rate produces a broader Na plateau that
 extends a factor of two further above the O-depleted region than when
 using the lower limits.
Mixing into this region can create large Na enhancements without affecting
 the O abundance.
This could further explain the results of Norris \& Da Costa 
 (\markcite{r46}1995a) and Briley et al. (\markcite{r122}1997).

The most important rates to understand in the NeNa cycle for determining
 the Na and Mg abundances at lower metallicities are the cycling reaction, 
 $^{23}$Na(${\it p},{\alpha})^{20}$Ne, and the leakage reaction,
 $^{23}$Na(${\it p},{\gamma})^{24}$Mg.
The former reaction contains an upper limit on the 36 keV resonance,
 which can become important for lower temperatures around $T_{9} =$ 0.04.
At temperatures higher than this, the contribution quickly becomes insignificant
 when compared to the other well-known terms.
The leakage reaction has an uncertain resonance near 137 keV which
 becomes important only near $T_{9}$ = 0.07, which is never attained in
 our models, even near the tip of the RGB.
Using these rates, we determine the ratio of leakage to cycling, R$_{lc}$,
 and compare it with the same ratio from CF88.
The results are given in Figure 22, where the curve represents the ratio of
 the rates given by EC95 over the ratio of the rates given by CF88.
\begin{figure}
\plotone{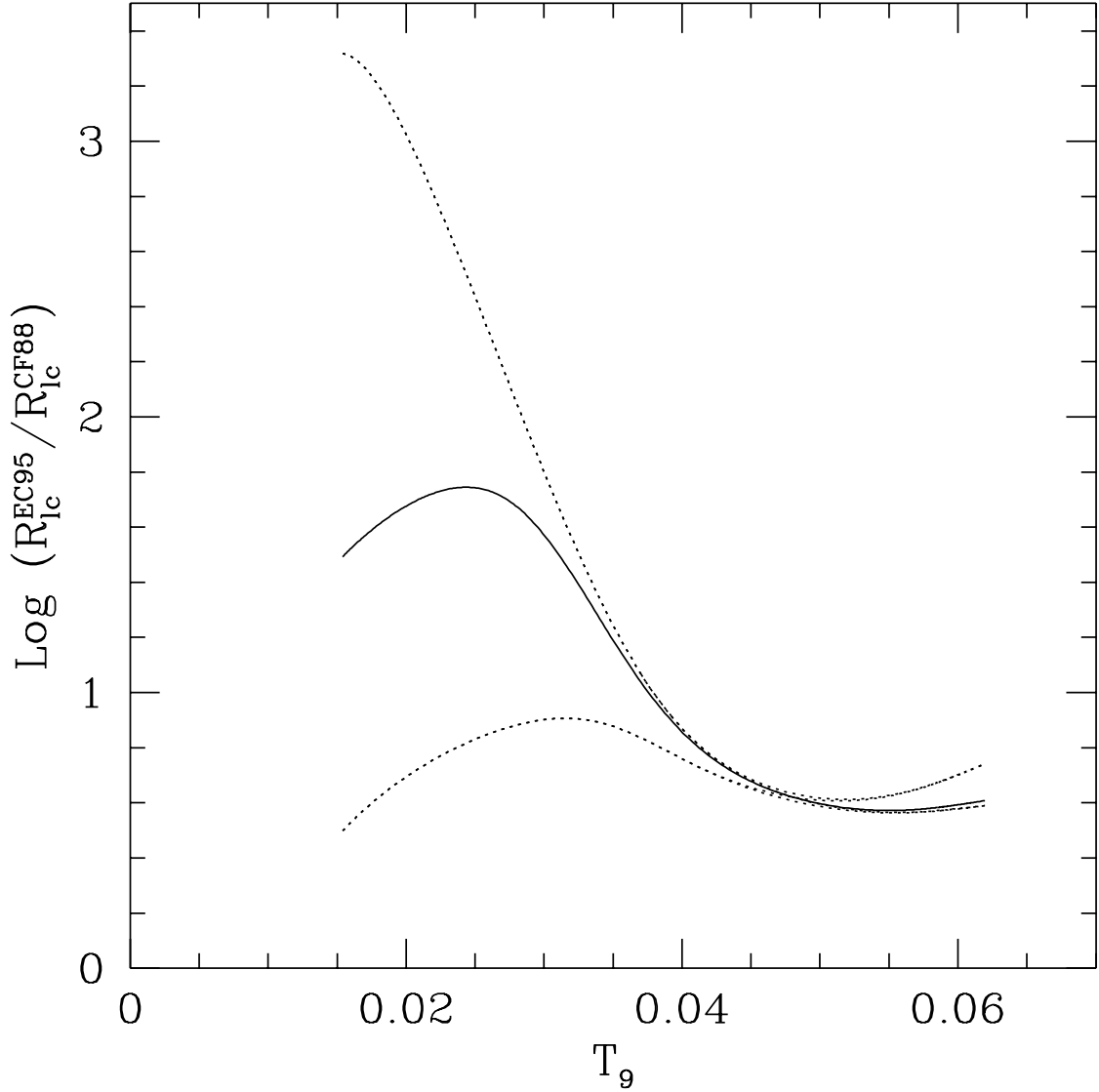}
\caption{The ratio of the ``leakage-to-cycling'' ratios
 (R$_{lc}$) of the EC95 and CF88 rates.  The graph is read, for example, by
 saying that at $T_9$ = 0.035, the EC95 rates predict 10 times as much leakage
 from the NeNa cycle as the CF88 rates. The dotted lines represent the 
 published uncertainties in the EC95 rates, while the solid lines represents
 the EC95 rates used in this work.}
\end{figure}
At low temperatures, R$_{lc}$ is nearly 55 times stronger with the new rates,
 but only 4-6 times stronger in the important temperature range, 0.04 $\leq$ 
 $T_{9}$ $\leq$ 0.06.
Varying the values of the uncertain terms, from their least to highest 
 possible contributions, in all combinations has the effect of introducing a 
 factor of 10 uncertainty in R$_{lc}$ at $T_{9}$ = 0.03.
For $T_{9}$ = 0.05, the total range in uncertainty is only $\pm$ 1\%.
Although the comparative leakage rate can be much faster at lower temperatures,
 the absolute rate for the $^{23}$Na$({\it p},{\gamma})^{24}$Mg reaction
 is still about 5 orders of magnitude less at $T_{9}$ = 0.03 than at $T_{9}$
 = 0.05.
Therefore, the uncertainty in the leakage at lower temperatures has an 
 insignificant effect on the final abundances.
Finally, from Figure 22, it is apparent that the eventual production of
 $^{27}$Al through the connection between the NeNa and MgAl cycles at the shell
 temperatures becomes more likely with the rates of EC95.

\subsection{MgAl Cycle Reactions}

In this section we explore how the changes in the reaction rates of the MgAl
 cycle might affect the abundances of Mg and Al in general, and $^{24}$Mg
 in particular.
In order to match Shetrone's observations of M13 giants, the abundance of
 $^{24}$Mg must somehow be reduced, at least near the tip of the RGB.
To begin we explore how changes in the proton-capture rates of $^{23}$Na, 
 $^{25,26}$Mg, and $^{27}$Al might affect the production and destruction
 of $^{24}$Mg.
One method which might accomplish this is to slow the rate of leakage from
 the NeNa cycle.
Figure 23 shows the rate of destruction of $^{24}$Mg compared to its rate
 of production from $^{23}$Na and $^{27}$Al in the Z = 0.0001 helium-flash
 model, which has the largest range in temperature (see Figure 9).
\begin{figure}
\plotone{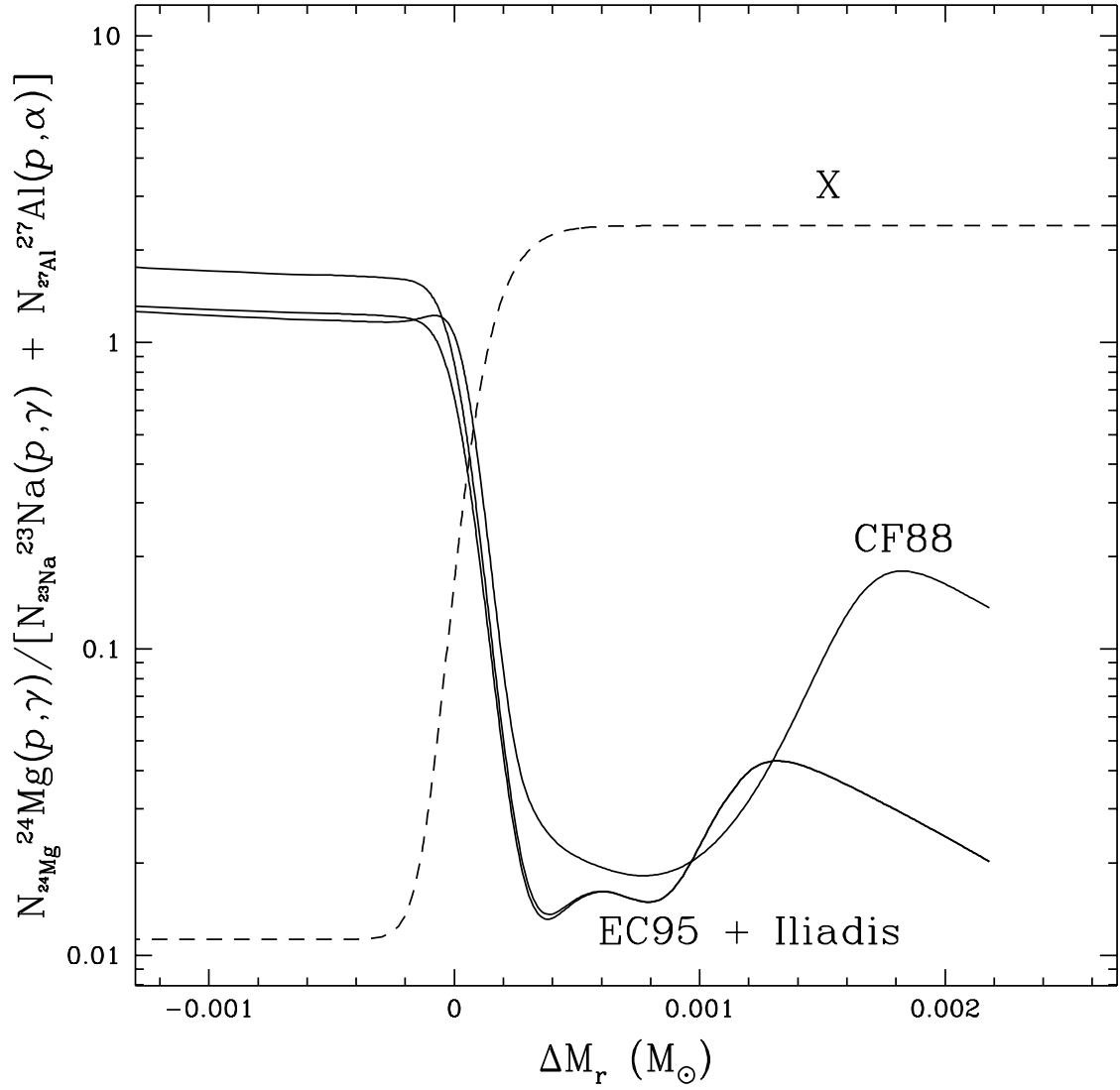}
\caption{A comparison of the destruction-to-production
 ratio of $^{24}$Mg by proton-capture nucleosynthesis for the rates of
 CF88 and newer results.}
\end{figure}
The new rates for the $^{27}$Al$({\it p},\alpha)^{24}$Mg reaction are from
 Iliadis (private communication; see below).
One curve shows the destruction-to-production ratio for the CF88 rates
 and two show the same for the upper and lower limits of the EC95 rates.
Since this is done for an actual model, the number of available reactant 
 nuclei is taken into account.
If we compared the production rates of $^{24}$Mg only from the $^{23}$Na
 proton captures and ignored its production from 
 $^{27}$Al$({\it p,{\alpha}})^{24}$Mg, then the CF88 curve would be
 nearly a factor of 6 greater than the EC95 curves; 
 however, because the $\alpha$-decay reaction from $^{27}$Al 
 is more rapid with the CF88 rates than with the newer Iliadis rates,
 all three ratios are similar.
Above the H shell, where the destruction-to-production ratio is less than unity,
 it is apparent that the range in uncertainty due to
 the EC95 rates cannot weaken the leakage from the NeNa cycle enough
 to allow the destruction rate of $^{24}$Mg to overcome its production rate. 

Now let us put aside the quoted uncertainties from EC95 and explore some other 
 nuclear-physics limits.
We begin by completely shutting off the NeNa-cycle leakage.
This will allow the $^{24}$Mg to deplete with only marginal production from
 $^{27}$Al$({\it p,{\alpha}})^{24}$Mg at higher temperatures.
Figure 24 shows Mg plotted against Al for both leakage via the EC95 rates
 and no leakage, for the Z = 0.0004 sequence at the RGB tip.
\begin{figure}
\plotone{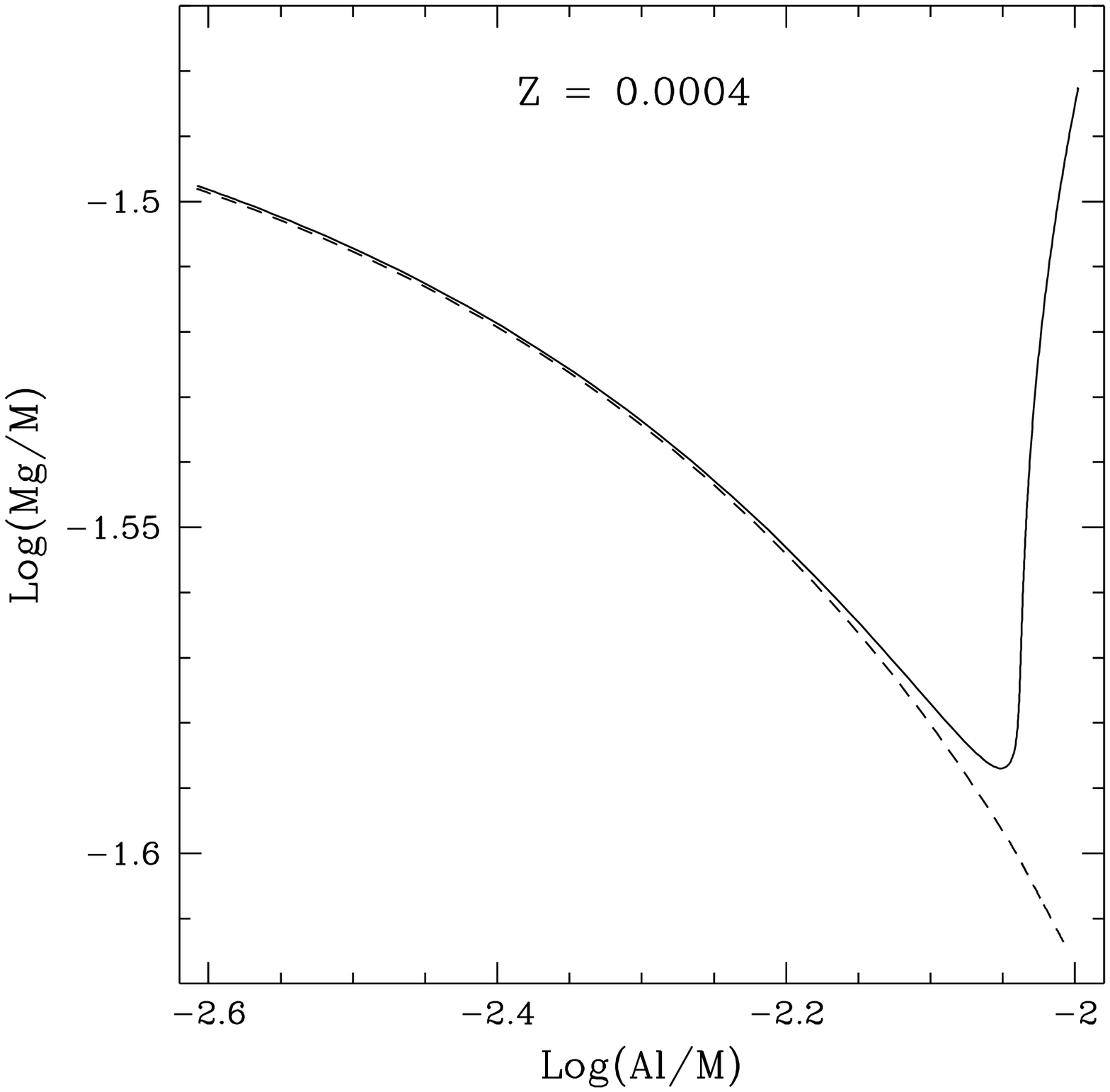}
\caption{Mg vs. Al at the RGB tip of the Z = 0.0004
 sequence for the EC95 rates chosen in this study (solid line) and for no
 leakage from the NeNa cycle (dashed line).}
\end{figure}
The initial anticorrelation at low Al/M is due to depletions in $^{25,26}$Mg.
We see from this figure that with NeNa-cycle leakage occurring,
 the $^{24}$Mg abundance increases at high Al causing the upturn in the 
 Mg/M abundance.
On the contrary, once the leakage is shut off, the total Mg abundance 
 diminishes, resulting in a total change of $\sim -$0.1 dex in Mg and $+$0.6
 dex in Al.
According to Figure 6 in Shetrone (\markcite{r116}1996b), the total Mg
 depletes by $\sim$ 0.4 dex while the Al increases by $\sim$ 1 dex.
Clearly, limiting the leakage from the NeNa cycle alone is not enough to
 account for the observations.
In addition, the extra-depletion of $^{24}$Mg with the leakage shut off 
 might occur too deep within the H shell to be mixed outward.

The $^{25}$Mg proton captures are from Iliadis et al. (\markcite{r110}1996)
 and are well-determined in the range of $T_9$ from 0.01 to 0.06, with 
 the largest uncertainty being less than a factor of two at $T_9$ = 0.06.
For the $^{26}$Mg proton capture reactions, we used the average values
 from the tables given by Iliadis et al. (\markcite{r109}1990).
The range in uncertainty does not become significant until $T_{9} \sim$ 0.05
 where it spans a factor of $\sim$ 100.
The $^{25,26}$Mg$({\it p,{\gamma}})$ reactions serve to initially enhance
 the $^{27}$Al abundance at cooler temperatures and ultimately to bridge
 the reactions between $^{24}$Mg and $^{27}$Al at higher temperatures.
Figure 25 shows the effect of varying the $^{26}$Mg proton capture rate between
 its upper and lower limits for the Z = 0.0004 sequence.
\begin{figure}
\plotone{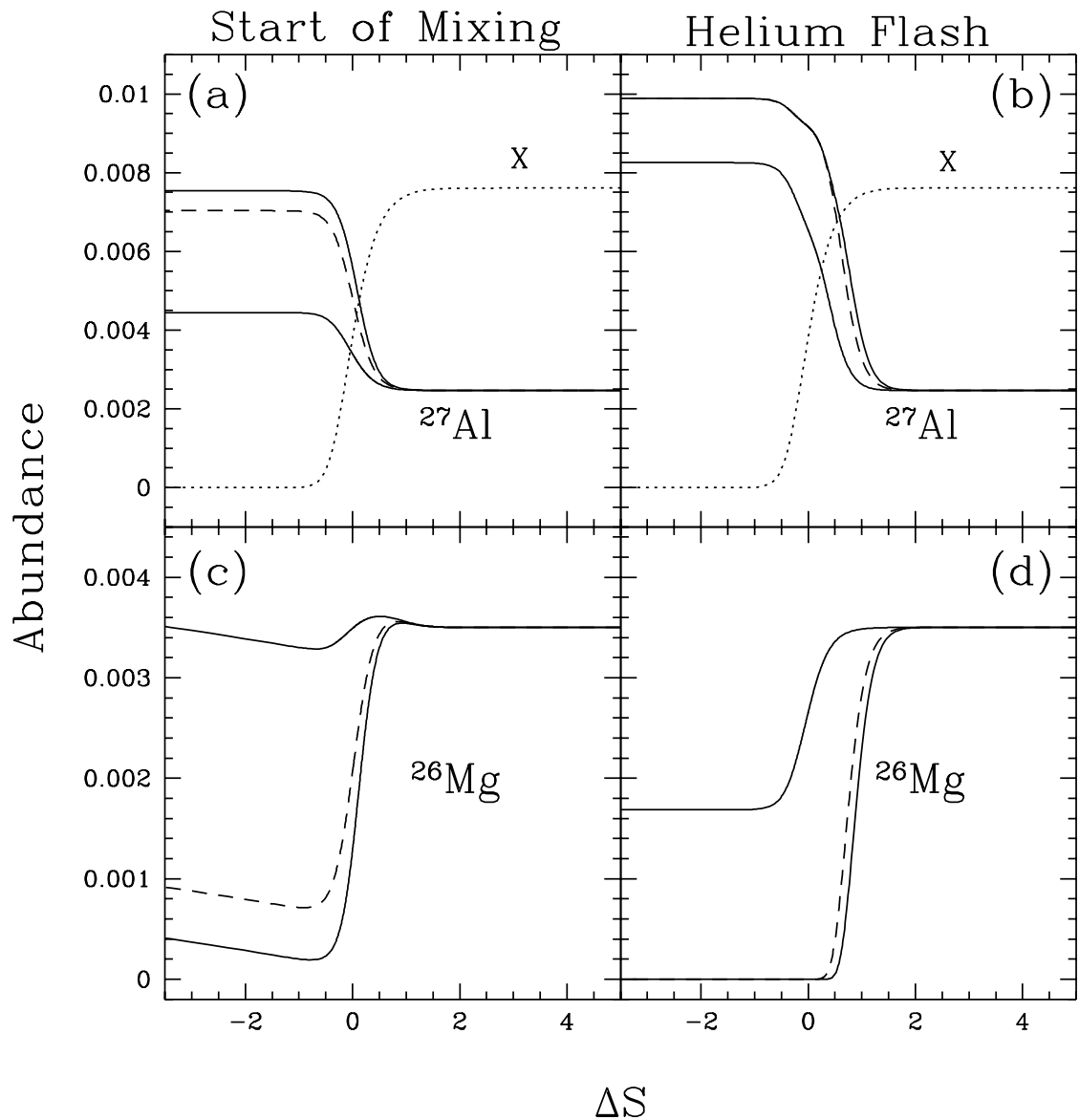}
\caption{The synthesis of $^{27}$Al (top panels) and
 $^{26}$Mg (bottom panels) at the start of mixing (left panels) and at the RGB
 tip (right panels) for the Z = 0.0004 sequence. The solid lines represent the 
 range of abundances obtained from the upper and lower limits, the dashed
 line is the result of using the average of the upper and lower limits of the
 ${26}$Mg proton-capture rate, and
 the dotted line is the H-mass fraction scaled by 100.}
\end{figure}
The average of the limits is about the same as the CF88 rates scaled 
 downward by a factor of $\sim$ 16 at $T_9$ = 0.02 to a factor of $\sim$ 4
 near $T_9$ = 0.05.
The difference in the resultant abundances is larger on the lower RGB when 
 the H shell is approaching the temperature regime where the difference
 between the upper and lower limits of the rates becomes significant.
Using the upper limits creates 0.22 dex more $^{27}$Al at the center of the
 H shell at the start of mixing in comparison to using the lower limits.
The difference in the location around the H shell of where the elements
 change abundance is more pronounced on the upper RGB, where the now very
 different limits manifest their sensitivity to the higher temperatures.
For example, toward the RGB tip, $^{27}$Al becomes enhanced farther above the
 H shell when the upper limit is used than when the lower limit is used.
The result is that the upper limit produces 0.16 dex more $^{27}$Al at the 
 top of the H shell (${\Delta}$S = $+$1) at the tip of the RGB. 
The consequences of both the difference in the production rate and the point
 of production above the H shell depend on the depth of mixing.

The overall effect of the uncertainty in the $^{26}$Mg proton-capture reaction
 rate is seen in all four sequences.
As the metallicity decreases, the uncertainty becomes less important because 
 the temperatures are high enough that the $^{26}$Mg is diminished near the top
 of the H shell for all but the lowest rates.
In these cases, all that is necessary to create large overabundances of 
 $^{27}$Al is that the $^{26}$Mg proton capture proceed more rapidly than
 the $^{24}$Mg proton capture, which it does.
This rate will, however, affect the observed Mg isotope ratios, as will the
 $^{25}$Mg$({\it p,{\gamma}})^{26}$Al$({\beta}^{+})$ reactions.

Lastly, we explore whether the final two reactions in the MgAl cycle might have
 important consequences for the abundance of $^{24}$Mg.
These reactions are the alpha- and gamma-decays of the $^{27}$Al proton
 capture.
They determine the cycling nature of the MgAl cycle just as the
 $^{23}$Na $(\it p, {\alpha})$ and $(\it p,{\gamma})$ reactions do the same
 for the NeNa cycle.
In Paper I we showed how the CF88 rates promote rapid cycling of $^{27}$Al
 back into $^{24}$Mg.
However, contemporary works of CF88 disputed MgAl cycling at low 
 temperatures and showed the MgAl reactions did, in fact,
 form a chain (Champagne et al. \markcite{r111}1988; Timmermann et al.
 (\markcite{r112}1988).
In this work, we use the rates generously provided to us by Dr. Christian
 Iliadis, for both the alpha- and gamma-decay reactions
\footnote{Dr. Iliadis includes the following comments with his rates:
The cycling-to-leakage ratio, which we denote as R$_{cl}$, probably depends
 mostly on the contributions of the E$_R$ = 72 keV and 85 keV (where E$_R$ is 
 the resonance energy) resonances in the $^{28}$Si nucleus.
If the reactions proceed only through the 85 keV resonance, then R$_{cl}$ is
 $\sim$ 0.66.
If the contribution from the 72 keV resonance is significant, then R$_{cl}$ is
 closer to 4, as shown in Figure 26b.
The ratio is better determined than the absolute values of the rates due
 to the difficulties in measuring very small ( $\sim 10^{-15}$ eV) proton
 partial widths at the low energies found in the H shell.
He estimates that the uncertainties in the absolute magnitudes of the rates
 are at least a factor of ten.}.
\begin{figure}
\plotone{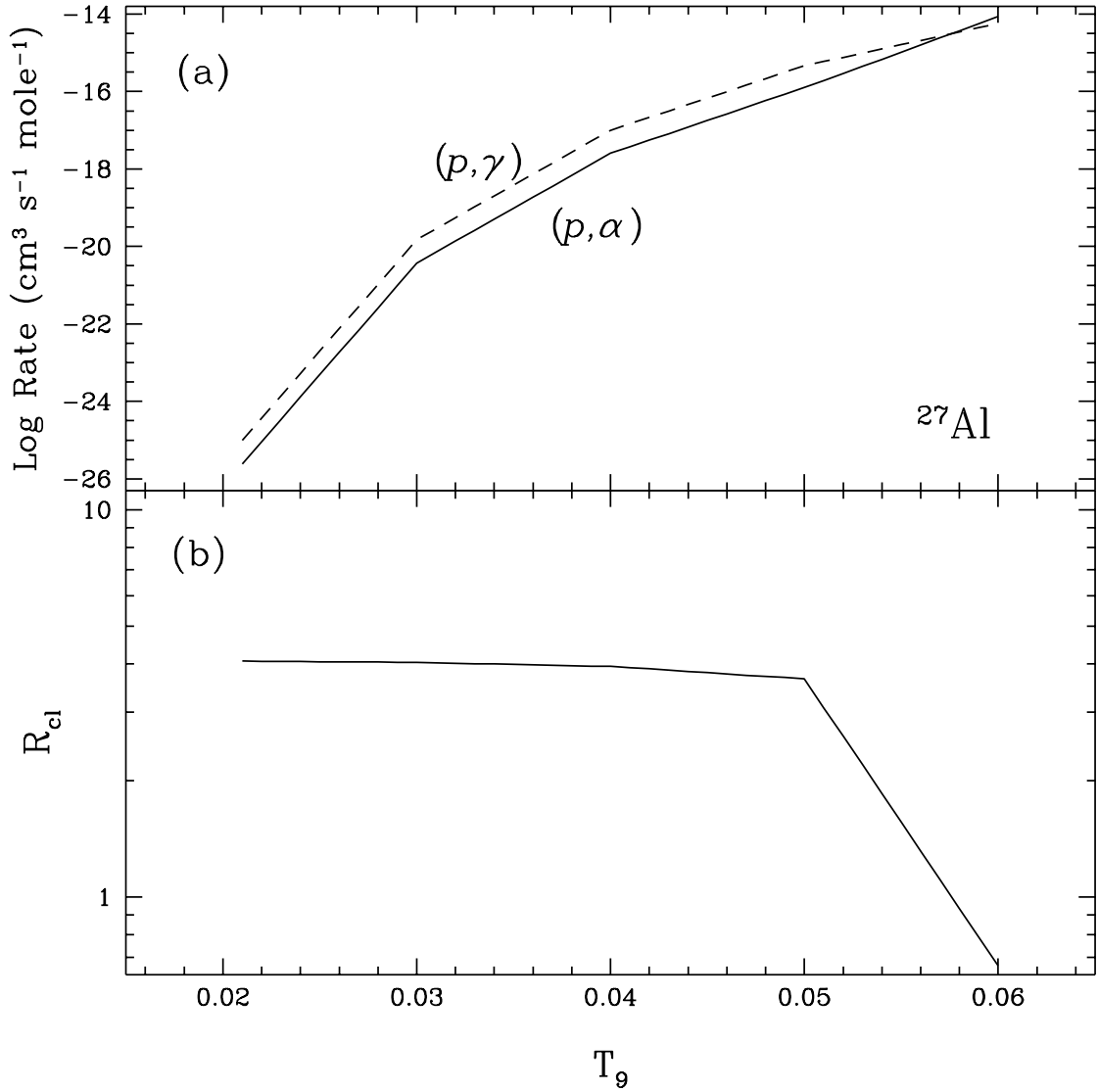}
\caption{(a) The $^{27}$Al proton capture rates from Iliadis (1996,
 private communication).  (b) R$_{\rm cl}$, the ${\alpha}/{\gamma}$-decay
 ratio from the rates in panel (a).}
\end{figure}
Figure 26 shows the two rates on an absolute scale and their ratio
 for $T_9$ = 0.02 - 0.06.
The top panel indicates that these two rates proceed very slowly at lower
 temperatures when compared to the lifetime of the RGB.
Even at higher temperatures, the destructive proton captures cannot keep
 pace with the creation of $^{27}$Al from the other reactions in the cycle.
The bottom panel shows that the cycling-to-leakage ratio, R$_{\rm cl}$,
 for the MgAl cycle is weak, even with the upper limits for the alpha decay
 and the lower limits for the photon decay.

We examine the importance of the cycling reaction on both the $^{24}$Mg
 and $^{27}$Al abundance by setting the $({\it p,\alpha})$ reaction equal
 to zero.
\begin{figure}
\plotone{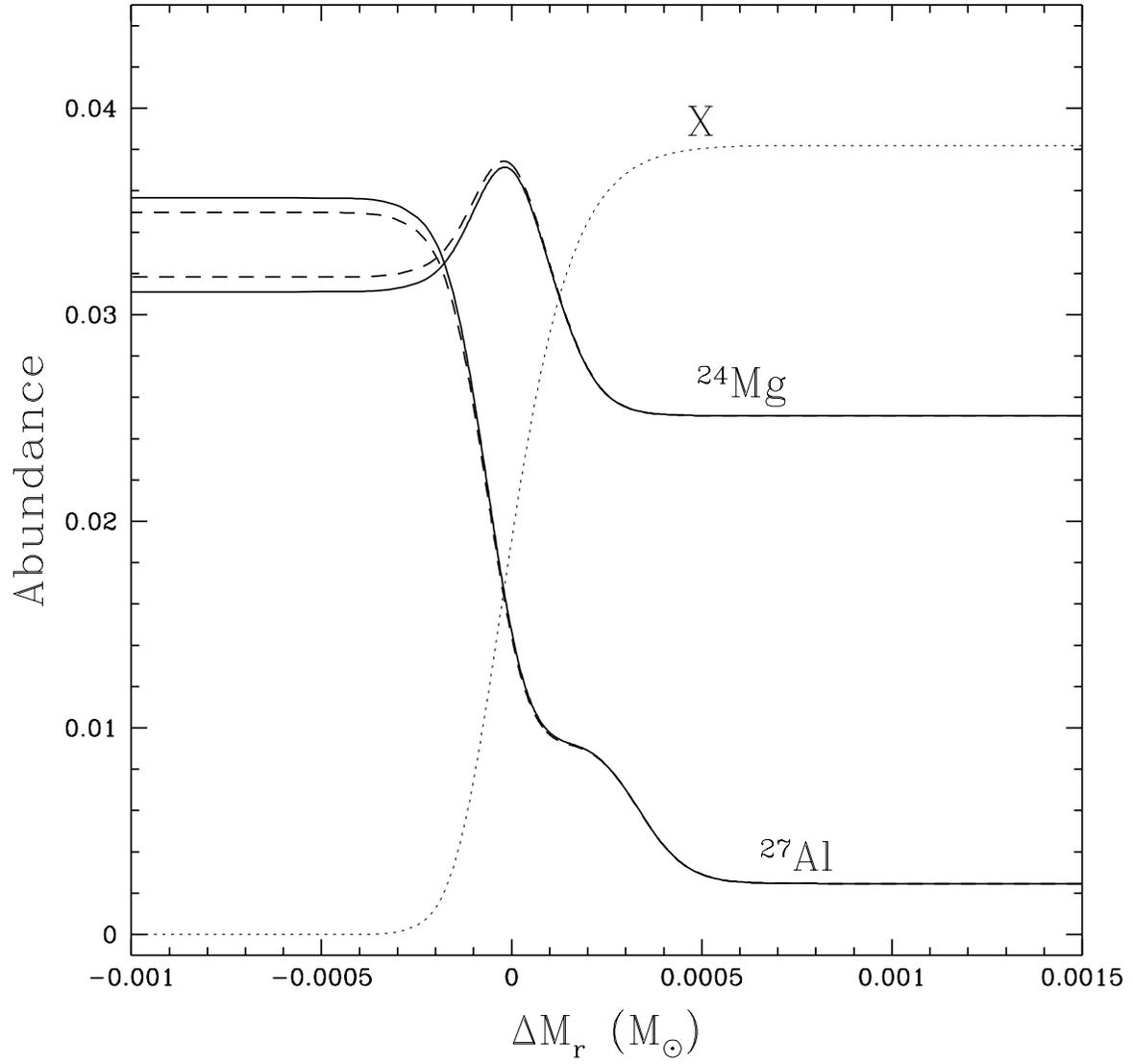}
\caption{The $^{24}$Mg and $^{27}$Al yields with cycling 
 (dashed line) and no cycling (solid line). The dotted line is the H-mass
  fraction scaled by a factor of 20.}
\end{figure}
The results are shown in Figure 27 for the model with the highest shell 
 temperature, i.e., the Z = 0.0001 sequence at the onset of the helium flash.
Clearly the difference in the $^{24}$Mg abundance is negligible; there is
 only a 0.7\% reduction at the center of the shell.
As a result of no cycling, the $^{27}$Al abundance increases by only 1.8\%.
Even at these hot temperatures, the destruction of $^{24}$Mg is not affected by
 the reactions which complete the MgAl cycle.
If its abundance is to be reduced to match the observations in M13
 (whose stars  theoretically never attain the high temperatures suggested 
 by this model), then another nuclear reaction-rate scenario is needed.
We therefore conclude that the overall destruction of $^{24}$Mg does not
 appear to be possible above the center of the H shell according to our present
 knowledge of the nuclear reaction rates and stellar interiors.

Since (1) totally depleting the $^{25,26}$Mg, (2) shutting off the leakage
 from the NeNa cycle, and (3) limiting the cycling in the MgAl cycle do not
 cause a sufficient enough depletion of $^{24}$Mg and enhancement of $^{27}$Al
 above the H shell to match the observations of the brightest M13 giants, we
 look for an alternative solution.
One possibility is presented by Langer, Hoffman, \& Zaidins
 (\markcite{r118}1996) who suggest that increasing the shell temperature
 to $T_{9} \sim$ 0.07 would reproduce these observations.
They rely on thermal instabilities in the H-shell to produce the non-canonical
 temperatures.
However, it has been shown (Von Rudloff, VandenBerg, \& Hartwick
 \markcite{r126}1988) that the H shell is stable against these thermal 
 perturbations.
In addition, by not using evolutionary sequences, they are not able to follow
 the effect of this extreme temperature on the evolutionary track of
 a star.
A temperature of $T_{9}$ = 0.07 would require the hottest model for the
 Z = 0.0004 sequence to increase its temperature by nearly 1.4 x 10$^{7}$ K.
This would result in a $\sim$ 1.6 dex increase in the luminosity of the RGB
 tip for this sequence.
Thus, we rule out their conclusion as inconsistent with the observed
 color-magnitude diagrams of globular clusters.

Instead, we explore what changes in the $^{24}$Mg proton-capture rate would be
 sufficient to match these limited observations.
Zaidins \& Langer ({\markcite{r125}1997) suggest that theoretically, there is
 room for a factor of 35 increase in the $^{24}$Mg proton-capture rate based
 on a poorly known width of the lowest resonance at 2468 keV in $^{25}$Al.
In Figure 28 we present the effect of increasing this rate by various
 amounts on the one model which most closely resembles the brightest M13
 giants.
The top panel demonstrates that in order to reduce the total Mg abundance by
 0.4 dex to match the observations, the $^{24}$Mg proton-capture rate would
 need to be increased by at least a factor of 35, assuming that all the
 other reactions which produce $^{24}$Mg are unchanged.
The corresponding increase in [Al/M] is about 1.1 dex, which agrees well
 with the observations.
However, this requires that mixing penetrates to the base of the H shell.
The resultant Al abundance from $^{24}$Mg is less sensitive to the $^{24}$Mg 
 proton-capture rate than the $^{24}$Mg abundance; there is approximately a
 0.1 dex difference in the $^{27}$Al production for each rate at 
 the base of the H shell.
At a mixing depth of say 20\% H depletion, the change in the total Mg
 abundance for the factor of 50 increase in the $^{24}$Mg proton-capture
 rate is not distinguishable from having no increase in the rate.
The questions which remain to be answered are (1) how much of an increase in the
 $^{24}$Mg$({\it p,{\gamma}})^{25}$Al reaction is possible, and (2) how
 deeply into the H shell can the mixing realistically penetrate?
At this point the best we can say about the first question is that the rate
 probably needs to be substantially increased in the H shell temperature range.
Whether this can happen is currently being considered (Iliadis, private
 communication).
The second question will be explored in better detail once mixing is
 studied in these sequences.
\begin{figure}
\plotone{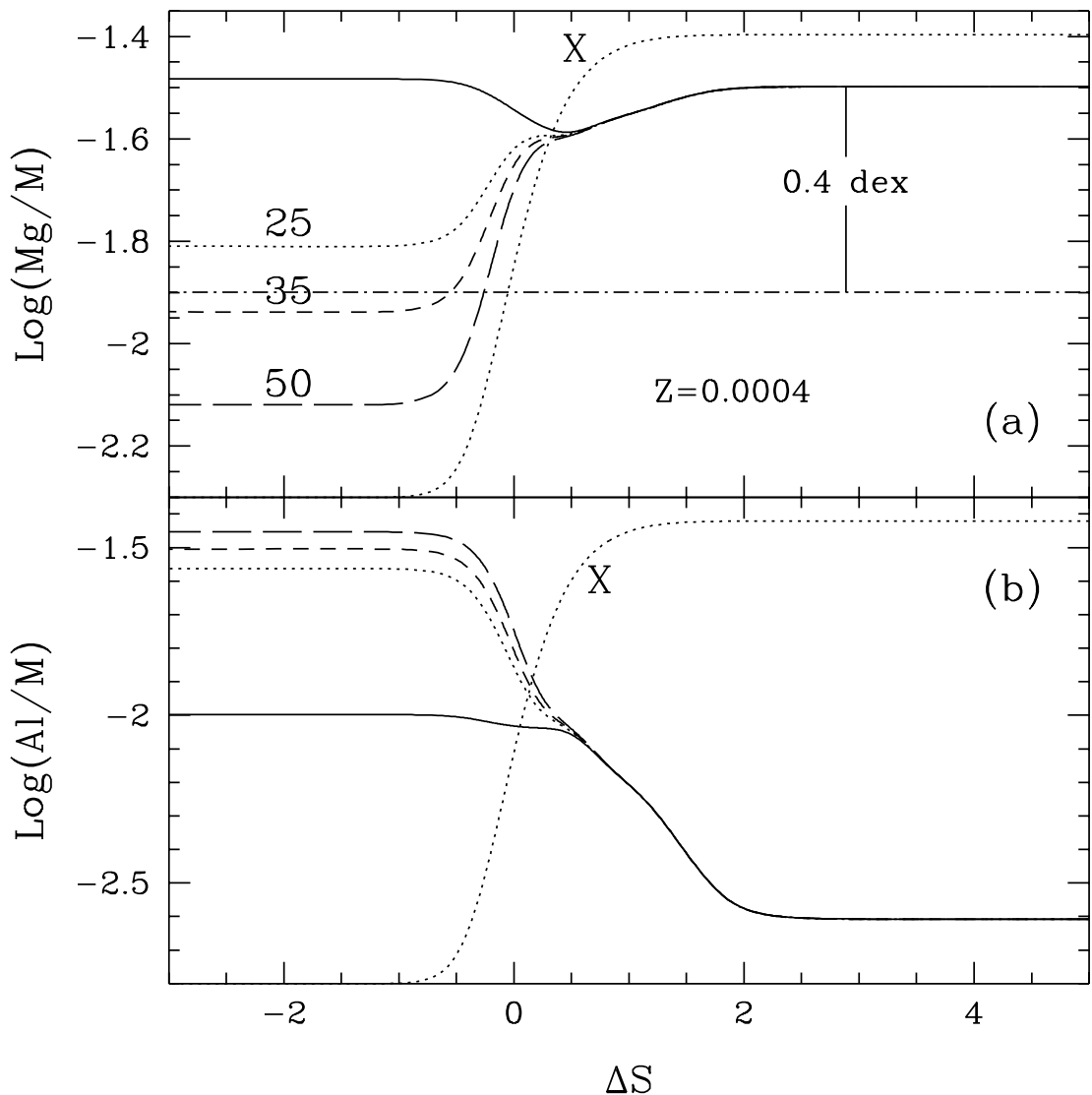}
\caption{The change in the total Mg (top panel) and Al
 (bottom panel) abundances in the RGB tip model of the Z = 0.0004 sequence
 as a consequence of increasing the $^{24}$Mg$({\it p,{\gamma}})^{25}$Al 
 reaction rate of CF88 by a factor of 25, 35, and 50. The dotted line traces
 the H shell. The dot-dashed line in panel (a) indicates a depletion of
 0.4 dex in the total Mg abundance relative to the envelope value.}
\end{figure}

\section{Discussion}

We have combined a nuclear reaction network with detailed stellar
 models in order to follow the evolution of the
 light abundances around the H shell of globular cluster RGB stars.
In so doing, we have qualitatively reproduced many of the observations
 which show luminosity- and metallicity-dependent variations in the
 abundances of C, N, O, Na, Mg, and Al.
We conclude this work with a synopsis of the results with an eye towards
 further work which will incorporate a mixing algorithm into our calculations.

From the stellar physics, we conclude that both the metallicity and luminosity
 of a star play an important role in determining the surface abundances.
We have shown: (1), that  ${\tau}_{\rm shell}$ decreases with increasing 
 metallicity.
This allows less time for the nuclear burning to alter the internal abundances
 before the H shell overtakes the processing region.
(2), the more metal-rich sequences have steeper temperature gradients
 creating more narrow burning regions.
And (3), these sequences burn at lower temperatures around the H shell and
 therefore do not experience all the reactions that the lower-metallicity 
 sequences can.
These three points lead to the conclusion that the low-metallicity sequences
 should experience greater variations in their surface abundances in more
 elements than the high-metallicity sequences, assuming mixing does occur.

Within a given sequence, the stellar physics leads to similar conclusions.
As a star evolves, ${\tau}_{\rm shell}$ decreases and the processing 
 regions contract.
This, by itself, would tend to reduce the amount of matter that can be mixed.
Competing with this in the higher-luminosity models is a higher temperature
 which can cause more matter to be processed on a shorter timescale.
In addition, the higher temperatures of the brighter models cause some
 nuclear reactions to occur that were not possible on the lower RGB.
Thus, we expect to see luminosity dependent variations within a given
 sequence, with the extent depending heavily on the mixing parameters.

As our models show no changes in the abundances of any 
 element heavier than $^{27}$Al, we discuss the changes one might expect to 
 see in the CNO-, NeNa-, and MgAl-cycle isotopic abundances
 as a function of luminosity and metallicity.
We have shown that there exists two separate regions above the H shell in
 which $^{12}$C and $^{16}$O are depleted at all luminosities and 
 metallicities.
Therefore, stars which undergo mixing on the RGB should exhibit a C vs. N
 and perhaps a O vs. N anticorrelation regardless of metallicity.
The mixing parameters could prevent the products of the O-N conversion from
 being brought to the surface in some higher-metallicity stars where the O-N 
 processed region is small compared to the C-depleted region.
The one ubiquitous conclusion which we can make is that, irrespective of
 metallicity, any star which shows C variations due to mixing should also
 exhibit a low $^{12}$C/$^{13}$C ratio.
In addition, any star which shows O variations due to mixing should also
 show a $^{16}$O/$^{17}$O ratio near its equilibrium value, or slightly
 smaller due to the immediate increase in $^{17}$O from $^{16}$O at the onset
 of O depletion.
According to the new $^{17}$O proton-capture rates of Blackmon et al.
 (\markcite{r106}1995) this value is between 50 and 150 and should
 peak at brighter luminosities.
Currently, there are no O isotope data for globular clusters.
Perhaps observations of this quantity in stars which show large O depletions
 that would indicate mixing into the O equilibrium region, such as the 
 super O-poor stars in M13 seen by KSLS93, can help constrain these rates.
We submit that the $^{17}$O abundance might best be seen from the
 $^{12}$C$^{17}$O IR system at 2.3 {\micron} and 4.6 {\micron} (Balachandran,
 private communication).

As for the NeNa cycle abundances, we conclude that all the sequences are
 capable of showing some Na enhancements as a result of a single proton
 capture on $^{22}$Ne.
This occurs nearer to the surface than the O-depleted region for lower 
 luminosity models independent of metallicity and eventually overlaps the
 O shell as the sequence evolves.
Depending on the mixing depth, this can result in a slight increase in $^{23}$Na
 with little variation in O.
Slightly deeper mixing, especially in brighter models, should cause an
 anticorrelation between Na and O.
Extra-Na enhancements from $^{20}$Ne are seen inside the O-depleted region
 when mixing begins for the Z = 0.0001 and Z = 0.0004 sequences.
The Z = 0.004 sequence shows larger Na enhancements from the NeNa cycle
 only after the onset of mixing, and the Z = Z$_\odot$ sequence exhibits
 hardly any NeNa-cycle processing.
Thus, we anticipate that the lower-metallicity stars might exhibit large
 Na enhancements if they also show large O depletions.
Higher-metallicity stars, however, which show large O depletions at lower
 luminosities might not have large Na enhancements until they evolve further
 up the RGB.
Even then, the mixing would have to be very deep (near the center of the
 H shell) and would have to compete with a decreasing ${\tau}_{\rm shell}$.

We also explore the cycling nature of the NeNa cycle using the new rates
 of EC95 compared to the more widely used rates of CF88.
We find that the EC95 rates promote leakage from the cycle at four times the
 rate promoted by CF88 for 0.04 ${\lesssim}\ T_9\ {\lesssim}$ 0.06, with little
 variation caused by the uncertain resonances in the $({\it p,{\gamma}})$
 and $({\it p,{\alpha}})$ reactions.
This causes an increase in the $^{24}$Mg abundance at higher temperatures 
 despite it being depleted to create $^{27}$Al.

Finally, we discuss the MgAl cycle elements.
These change only inside the O-depleted region and should be correlated
 with a diminution of O and an enhancement of N.
The abundance yields for these elements are very sensitive to temperature as
 manifested by both luminosity and metallicity.
To begin, $^{27}$Al is produced by all three Mg isotopes.
In the Z = Z$_\odot$ sequence, it comes solely from proton captures on
 $^{26}$Mg and only near the tip of the RGB.
For the same sequence, the meta-stable isotope $^{26}$Al is produced directly
 from $^{25}$Mg, and decays into $^{26}$Mg as it is overtaken
 by the H shell.
Assuming sufficiently quick (${\tau_{\rm mix}}$ less than the decay lifetime
 of $^{26}$Al) and deep (${\Delta}$S $\sim$ 0) mixing for such a sequence, we
 would anticipate a large depletion of the $^{25}$Mg abundance, a minor
 depletion of the $^{26}$Mg abundance that is correlated with a slight
 enhancement of $^{27}$Al, and no change in $^{24}$Mg at all RGB luminosities.
However, if the mixing is slower, we would expect the $^{25}$Mg abundance
 to be depleted completely into $^{26}$Mg, resulting in a possible
 enhancement of $^{26}$Mg despite the slight increase in $^{27}$Al.
As the metallicity decreases to Z = 0.004, the $^{27}$Al production
 starts to come more from both $^{25}$Mg and $^{26}$Mg.
Again though, there is no change in the $^{24}$Mg abundance.

We have shown that we are sensitive to the $^{26}$Mg proton-capture 
 rate above the top of the H shell for all metallicities.
Both the upper and lower limits produce a significant amount of $^{27}$Al, but
 at slightly different depths.
Thus, $^{26}$Mg variations depend strongly on the depth of mixing and further 
 quantitative analysis shall have to wait until mixing is invoked.

The two lower-metallicity sequences show the largest changes in the $^{27}$Al
 abundance.
Due in part to the relatively large value of ${\tau}_{\rm shell}$, all the 
 $^{25}$Mg and $^{26}$Mg is converted into $^{27}$Al even at the
 start-of-mixing luminosity.
As the sequence evolves, the $^{27}$Al abundance increases dramatically 
 within the H shell from proton captures on $^{24}$Mg in the MgAl cycle.
Not all of this extra Al is likely to be reached by the mixing currents
 since it occurs in a region of large H depletion.
Thus, the Al variation is very dependent on the mixing parameters.
At this point, we can only conclude that lower metallicity sequences
 (Z $\lesssim$ 0.0004) should exhibit large Al enhancements that should
 correlate with a diminution of $^{25,26}$Mg and perhaps $^{24}$Mg.

The most difficult puzzle to solve is that of the observed $^{24}$Mg
 abundances in the M13 RGB and NGC 6752 tip stars.
According to our sequence at the relevant metallicity, we would expect no 
 depletion of $^{24}$Mg in the same region of Al enhancements above the
 center of the H shell, despite $^{24}$Mg being the source of the
 Al over-production.
This is because the EC95 rates provide sufficient leakage from the NeNa
 cycle into $^{24}$Mg to more than account for its depletion.
We showed in Paper I that even the ``standard'' CF88 rates predict a marginal
 increase in the $^{24}$Mg abundance due to leakage.
To reconcile our models with the observations we have tried limiting the 
 NeNa leakage by both taking its published lower limit and by shutting it
 off completely.
We have also tried stopping the cycling reaction in the MgAl cycle.
None of these procedures had much impact on the $^{24}$Mg abundance at
 realistic H depletions.
In order to decrease the $^{24}$Mg abundance by the observed amount, it
 is necessary to increase significantly the $^{24}$Mg proton-capture rate
 and to mix deeply into the H shell.
Finally, we rule out as a possible solution any dramatic increases in the
 shell temperature since this would contradict the observed RGB tip luminosity.
We are thus left in a quandary about how to approach the $^{24}$Mg anomalies.
This is clearly a problem whose resolution requires further observations,
 investigation into the $^{24}$Mg proton-capture rate, and the addition of 
 mixing into the sequences.

Putting the $^{24}$Mg uncertainties aside, we note that the proximity of the 
 large Al enhancements to the H shell and the observational
 evidence which supports mixing to this depth would indicate that some fraction
 of He must also be mixed to the surface.
If the extra mixing of He is a contributing factor of the blue horizontal 
 branches (HB) in some clusters as suggested by Langer \& Hoffman
 (\markcite{r95}1995) and verified though extensive modeling by Sweigart
 (\markcite{r115,r127}1997a,b), then one would expect these ``second-parameter''
 clusters to be super-abundant in Al.
Indeed, M13 (Kraft et al. \markcite{r113}1997) and NGC 6752 (Shetrone
 \markcite{r124}1997) fit this criteria .
This prediction can be further tested by searching for Al in clusters with
 similar metallicity to these clusters but with redder HB's, such as M3 and M5.
In fact, both M3 and M5 show less severe O depletions than M13 (Sneden et
 al. \markcite{r85}1992; KSLP92), which would indicate less severe mixing.
If the Al is anticorrelated with O, then M3 and M5 should not have the large
 Al enhancements or the $^{24}$Mg depletions seen in M13.

Pilachowski \& Sneden (\markcite{r72}1983) observed that NGC 288 has a very
 blue HB for its metallicity ([Fe/H] = -1.0).
This might indicate the presence of deep mixing.
However, based on our results, a cluster this metal-rich would not be
 expected to over-produce Al because it never attains high enough temperatures
 to burn $^{24}$Mg.
Thus, the most one might expect to observe in NGC 288 if deep mixing is 
 bringing He up to the surface to create the blue HB, is a slight increase
 in $^{27}$Al at the expense of $^{25,26}$Mg.

Future work will involve subjecting these sequences to a mixing algorithm 
 and the results will be reported elsewhere.
Having accomplished this, we will explore the abundances of
 other isotopes such as $^{3}$He and $^{7}$Li  which have significant
 cosmological implications.  (see, e.g., Spite \& Spite \markcite{r120}1982; 
 Charbonnel \markcite{r121}1996 )

\acknowledgements
We wish to thank Dr. David Arnett for allowing us the use of his nuclear
 reaction network.
We express our gratitude and indebtedness to Dr. Christian Iliadis for his
 thoughtful discussions of our nuclear reaction network and for his
 encouragement of our study of the effects of the uncertainties in the
 rates on our results.
It is also our pleasure to thank the members of Triangle Universities National
 Laboratory, particularly Ms. Denise Powell and Mr. Steve Hale, for their 
 careful experiments regarding the nuclear reaction rates in the NeNa and
 MgAl cycles.
We thank Dr. John Norris for his useful comments on our work prior
 to the writing of this manuscript and Dr. Bruce Carney for his insightful
 comments regarding Na and O abundances.
And for keeping us well informed of their latest observational data, Dr. 
 Michael Briley, Dr. Matthew Shetrone and Dr. Caty Pilachowski are
 gratefully acknowledged.
We also thank the referee for his/her careful reading of our manuscript and
 for the constructive comments which helped form the final draft of this work.
The work of A.V.S. is funded in part by NASA grant NAG5-3028 and that of
 R.A.B. by NSF AST93-14931 and NASA grant NAG5-3028.
Finally, R.M.C. wishes to acknowledge the NASA Graduate Student
 Research Program for financial support of his research.

\end{document}